\documentclass[journal]{IEEEtran}

\IEEEoverridecommandlockouts    

\usepackage{enumitem}
\usepackage{graphicx}
\usepackage{algorithmic}
\usepackage{epsfig}
\usepackage{wrapfig}
\usepackage{amsmath,amsthm, amssymb}
\usepackage[]{algorithm2e}
\usepackage{multirow}
\usepackage{multicol}
\usepackage{tabularx}
\usepackage[dvipsnames]{xcolor}
\usepackage{comment}
\usepackage[utf8]{inputenc}

\usepackage{array}
\usepackage{color}
\usepackage{hyperref}
\usepackage{breqn}
\usepackage{cite}

\newcommand{\R}{\mathbb{R}}
\newcommand{\N}{\mathcal{N}}
\newcommand{\E}{\mathcal{E}}
\newcommand{\G}{\mathcal{G}}

\newcommand{\A}{\mathcal{A}}

\begin{document}
%
\title{Distributed Optimization using Reduced Network Equivalents for Radial Power Distribution Systems}

\author{Rabayet~Sadnan~\IEEEmembership{Student Member,~IEEE},    and~Anamika~Dubey,~\IEEEmembership{Member,~IEEE}
\vspace{-5pt}
\thanks{
Rabayet Sadnan and Anamika Dubey are with the Department of EECS, WSU, Pullman, WA-99164, USA; e-mail: rabayet.sadnan@wsu.edu}
}


\maketitle
\vspace{-3pt}
\begin{abstract}
The limitations of centralized optimization methods for power systems operation have led to the distributed computing paradigm, particularly in power distribution systems. The existing techniques reported in recent literature for solving distributed optimization problems are not viable for power distribution systems applications. The essential drawback remains a large number of required communication rounds, i.e., macro-iterations among the computing agents to solve one instance of the optimization problem; the typical number of macro-iterations are in the order of $10^2\sim 10^3$. In this paper, a new and scalable distributed optimization method based on Equivalent Network Approximation (ENApp) is proposed to solve optimal power flow (OPF) for a balanced radial distribution system. Specifically, the distribution system's radial topology is leveraged to reduce the decomposed systems into upstream and downstream network equivalents. The proposed innovations reduce the required number of macro-iterations/communication-rounds for convergence by order of magnitude. The approach is validated using IEEE 123-bus and IEEE 8500-node test systems and is shown to converge to the same optimal solution as obtained using an equivalent centralized OPF  (C-OPF) model.
\end{abstract}

\IEEEpeerreviewmaketitle

\vspace{-0.3cm}
\section*{Nomenclature}
\addcontentsline{toc}{section}{Nomenclature}
\begin{IEEEdescription}[\IEEEusemathlabelsep\IEEEsetlabelwidth{$V_1,V_2,V_3$}]
\item[\textit{Sets}]
\item[$\G = (\N ,\E)$] Directed graph for the distribution system
\item[$\N$] Set of all nodes 
\item[$\E$] Set of all distribution lines
\item[$\N_D$] Set of all DER nodes in the system
\item[$\N_L$] Set of all load nodes in the system
\item[$A_m$] Directed graph representing the $m^{th}$ area.
\item[$\N_m$] Set of all nodes in area $A_m$ 
\item[$\E_m$] Set of all distribution lines in area $A_m$ 
\item[$\A_R$] Set of all directed graphs defining all areas in a given distribution system. Here, $\A_R = \{A_1, A_2,...\}$
\item[$\mathcal{S}$] Set of operating constraints for the distribution system 
\item[$\mathcal{S}_m$] Set of operating constraints for the area $A_m$. 
\vspace{0.1cm}
\item[\textit{Functions}]
\item[$F$] Cost function for the centralized problem
\item[$\Phi_m$] Cost function for the distributed problem, $A_m$
\item[$\psi_1$] Function projecting the network equivalence from upstream to downstream areas.
\item[$\psi_2$] Function projecting the network equivalence from downstream to upstream areas.
\vspace{0.1cm}
\item[\textit{Variables}]
\item[$y_{1}$] Complicating variable representing voltage at the shared boundary bus.
\item[$y_{2}$] Complicating variable representing power flow at the shared boundary bus.
\item[$X$] Set of power flow and decision variables for the overall distribution system
\item[$X_m$] Set of power flow, local, and complicating variables for area $A_m$ 
\item[$x_m$] Set of local power flow and decision variables for area $A_m$, excluding complicating variables.
\item[$N_A=|\A_R|$] Total number of areas in the network
\item[$x_{ij}$] Line reactance for line $(i,j)$
\item[$r_{ij}$] Line resistance for ine $(i,j)$
\item[$V_i$] Voltage magnitude for node $i$
\item[$v_i = |V_i|^2$] Squared voltage magnitude for node $i$
\item[$l_{ij}$] Squared magnitude of the current flow in line $(i,j)$
\item[$P_{ij}$] Sending end active power for the line $(i,j)$
\item[$Q_{ij}$] Sending end reactive power for the line $(i,j)$
\item[$p_{L_j}$] Active load demand at the node $j$
\item[$q_{L_j}$] Reactive load demand at node $j$
\item[$p_{Dj}$] Active power generation by the DER at node $j$
\item[$q_{Dj}$] Reactive power generation/demand by the DER at node $j$
\item[$S_{DRj}$] kVA rating of the DER at node $j$
\item[$I_{ij}^{rated}$] Thermal rating for the line $(i,j)$
\end{IEEEdescription}
\section{Introduction}
\IEEEPARstart{T}{he} optimization of power systems operations, both at the transmission and distribution levels, is crucial for an efficient and reliable grid. Traditionally, the grid operations are centrally managed upon solving an optimal power flow (OPF) problem using centralized optimization techniques; the OPF problem is typically modeled as non-linear programming (NLP) problem \cite{momoh1999review1,castillo2013survey}. With the increase in the number of controllable devices, especially in power distribution systems, a centralized optimization paradigm suffers from computational challenges and is susceptible to a single-point failure \cite{molzahn2017survey}. This has led to an interest in distributed optimization methods that (1) reduce the computational requirements on a single decision-making unit by distributing the problem into several smaller sub-problems that are computationally simpler; (2) result in a decision-making paradigm with multiple interacting agents that is robust to single-point failures; and (3) relax the need for communication between the central controller and all connected controllable/non-controllable assets.

The existing literature includes several methods to formulate the OPF as a distributed optimization problem. In general, these methods adopt the traditional distributed optimization techniques such as Augmented Lagrangian \& Method of Multipliers (ALMM), Alternating Direction Method of Multipliers (ADMM), Auxiliary Problem Principle (APP), Predictor-Corrector Proximal Multiplier method (PCPM), Analytical Target Cascading (ATC), etc., to model a distributed optimal power flow problem (D-OPF) \cite{molzahn2017survey,boyd2011distributed}. For D-OPF applications, the existing literature prefers methods based on ADMM and APP due to their superior convergence properties \cite{molzahn2017survey}. Specific to radial distribution systems, several methods based on the ADMM technique have been lately proposed \cite{zheng2015fully,peng2016distributed}. A D-OPF formulation includes macro-iterations to exchange the variables among the decision-making agents and the micro-iterations to solve each agent's local subproblems. The macro-iterations dictate the communication systems' requirements while micro and macro-iterations together decide the time-of-convergence (ToC) for the algorithm. Unfortunately, both ADMM and APP algorithms applied to OPF take a large number of macro-iterations ($\ge 100$ iterations), i.e., the message-passing between agents, to converge for relatively small systems \cite{erseghe2014distributed, dall2013distributed }. While several recent methods attempted to improve the ToC for local problems (micro-iterations) \cite{qu2017harnessing,mhanna2018adaptive,peng2016distributed}, they still take hundreds if not thousands of macro-iterations to converge. Another popular approach is the APP algorithm that linearizes the penalty term in the augmented Lagrangian \cite{kim2000comparison}. While the APP-based distributed OPF methods are relatively faster and require sharing fewer variables, the number of macro-iterations (i.e., communication rounds among distributed agents) needed to reach convergence is of the same order as in ADMM \cite{lu2017fully}. 

In power distribution systems, having a large number of macro-iterations for a single step of the optimization problem is not preferred as it will lead to significant communication delays with the decision-making. Thus, a practical implementation of such algorithms requires a very fast communication system to reach a converged solution within a reasonable time. Deploying such a communication system is expensive, and thus, they are typically not available in power distribution systems. Besides requiring a large number of macro-iterations, ADMM and APP based methods applied to D-OPF pose several other limitations, including, but not limited to, parameter tuning and requiring at least one grid-forming DER in each area. Lately, to address some of these challenges, real-time feedback based online distributed algorithms have been explored in the related literature for network optimization \cite{ cavraro2017local,bernstein2019real,magnusson2020distributed}. Generally, these algorithms do not wait to optimize for a time-step but asymptotically arrive at an optimal decision over several steps of real-time decision-making. For {\color{black}example, authors developed distributed optimal reactive power control in \cite{cavraro2017local}, that can regulate the voltage and track the network-level objective}. However, they cannot achieve a global optimum solution \cite{cavraro2017local}. Further, with 16 DERs in IEEE 123 bus system, this approach takes around 100s of macro-iterations to converge, thus is slow in tracking the optimal solution \cite{cavraro2017local}. In \cite{bernstein2019real}, authors developed a unified approach for real-time optimization of distribution grids. However, this approach requires a central coordinator to update the dual variables. In \cite {magnusson2020distributed}, authors designed a distributed feedback control to address some of the issues with the asynchronous and delayed communications among smart agents; still, the algorithm takes hundreds of iterations to converge/track the optimal solution for a mid-size feeder. This raises further challenges to the performance of the algorithm for larger feeders, especially during the fast varying phenomenon. 
While distributed optimization methods are argued to be computationally efficient for larger systems, the existing state-of-the-art methods are not practically viable for power systems applications. Among several limitations, the primary drawback remains the requirement for a large number of macro-iterations requiring several hundred rounds of communication for information exchange among distributed agents to solve one instance of the optimization problem. Thus, implementing such algorithms requires expensive communication infrastructure with high bandwidth and low communication delays. Also, the existing methods need tuning of several parameters to achieve convergence. Some techniques require intelligence at each node of the system, making it challenging to scale for larger systems. The online distributed control algorithms also suffer from the same challenges of slow convergence, resulting in sub-optimal performance and slow tracking of the optimal solutions.  

To address these challenges, in this paper, we propose a new distributed optimization method to solve the distributed optimal power flow (D-OPF) problem based on reduced Equivalent Network Approximation (ENApp) for radial distribution systems. The novelty lies in the proposed decomposition approach for the centralized optimization problem that actively leverages the radial topology of the distribution systems. The proposed approach follows the primal decomposition method combined with the block-coordinate descent algorithm for distributed optimization. It does not require any central coordinator for optimization and only involves communication among the neighbors, resulting in a distributed algorithm as defined by the recent relevant literature \cite{molzahn2017survey}. Further, no parameter tuning is needed making the approach robust for any operating conditions. The key contributions of this work are:
{\color{black}
\begin{itemize}[leftmargin=*]
    \item We proposed a new distributed optimization approach to solve the D-OPF problems in radial power distribution systems. Specifically, the radial topology of the distribution system is leveraged in developing a computationally tractable and practically viable D-OPF algorithm. 
    The proposed D-OPF method converges to the solution of the same quality as C-OPF and takes a significantly fewer number of macro-iterations compared to the state-of-art distributed optimization methods;
    \item We provide a convergence analysis for the proposed D-OPF algorithm. Specifically, it is shown that the proposed D-OPF algorithm is guaranteed to converge at-most linearly at the boundary bus. Further, if the local subproblems converge and the centralized OPF admits a feasible solution, the proposed D-OPF method is guaranteed to converge.     
    \item Extensive validation of the proposed approach is presented using IEEE 123-bus and IEEE 8500-node test system. It is demonstrated that the proposed D-OPF algorithm converges to the same solution as the C-OPF problem. Thus, the solution obtained from the D-OPF algorithm is of the same quality as obtained via solving an equivalent C-OPF algorithm. Further, compared to state-of-the-art D-OPF techniques, for example, ADMM, the proposed D-OPF algorithm takes an order of magnitude fewer communication rounds among distributed agents (macro-iterations) to converge. The applicability of the proposed approach is also demonstrated for an unbalanced distribution feeder. 
\end{itemize}}


\section{Modeling \& Problem Formulation}
In this paper, $(\cdot)^T$ represents matrix transpose; $(\cdot)^*$ represents the complex-conjugate; $| . |$ symbolizes the cardinality of a discrete set or the absolute value of a number; $(\cdot)^k$ represents the $k^{th}$ iteration;
and $j = \sqrt{-1}$; $\overline{(.)}$ and $\underline{(.)}$  denotes the maximum and minimum limit of a given quantity. 

\vspace{-0.3cm}
\subsection{Network Modeling}
Let us consider a radial power distribution network with $n$ nodes. Let $\N$ be the set of all nodes in the system, i.e., $\N = \{1,2,…,n\}$. Let the directed graph $\G = (\N, \E)$ is used to represent the distribution system. Here $\E$ denotes the set of all distribution lines connecting the ordered pair of buses $(i,j)$ i.e., from node $i$ to node $j$. The distribution lines are modeled as series impedance, $r_{ij}+jx_{ij}$ for $\forall(i,j)\in \E$. Let, $\N_L$ represents the set of load nodes, and $\N_{D}$ represents the set of all DER nodes. For a radial topology, the cardinality of sets $\N$ and $\E$ is related as the following: $|\N| = |\E|+1$. Further, $|\N| \geq |\N_L| \geq |\N_{D}|$. {\color{black}Let, for node $j$, $i$ is the unique parent node, and $k$ be the set of all children nodes. Then, in $k: j\rightarrow k$, $k$ represents the set of children nodes for the node $j$. Next, we denote $v_j = |V_j|^2 = V_j{V_j}^*$ as the squared magnitude of voltage at node $j$  $(V_j)$. Likewise, let $l_{ij}$ be the squared magnitude of current flow in branch $(i,j)$.} 

The network is modeled using the branch flow equations \cite{baran1989optimal1} defined for each line $(i,j)\in \E$ and $\forall j \in \N$.

\vspace{-0.3cm}
\begin{small}
\begin{IEEEeqnarray}{C C}
\small
\IEEEyesnumber\label{eqModel_nonlin} \IEEEyessubnumber*
P_{ij}-r_{ij}l_{ij}-p_{L_j}+p_{Dj}= \sum_{k:j \rightarrow k} P_{jk}   \label{eqModel_nonlin1}\\
Q_{ij}-x_{ij}l_{ij}-q_{L_j}+q_{Dj}= \sum_{k:j \rightarrow k} Q_{jk} \label{eqModel_nonlin2}\\
v_j=v_i-2(r_{ij}P_{ij}+x_{ij}Q_{ij})+(r_{ij}^2+x_{ij}^2)l_{ij}\label{eqModel_nonlin3}\\
v_il_{ij} = P_{ij}^2+Q_{ij}^2 \label{eqModel_nonlin4}
\end{IEEEeqnarray}
\end{small}
\vspace{-0.3cm}

\noindent where, $P_{ij}, Q_{ij} \in \R$ is the sending-end active and reactive power flows for $(i,j)\in \E$. Complex power $S_{L_j} = p_{L_j}+jq_{L_j}$ is the load connected and $S_{D_j} = p_{Dj}+jq_{Dj}$ is the power output of DER connected at node $j$. Note that, $\forall j \not\in \N_{D}$, $S_{D_j} = 0$. Also, if $j\not\in \N_L$, then $S_{L_j} =0$. 

\vspace{-0.3cm}
\subsection{DER Modeling}
The DERs are modeled as balanced single-phase photovoltaic modules (PVs), interfaced with the distribution system using smart inverters capable of {\color{black}two}-quadrant operation. The active power generation for the DER is specified by $p_{Dj}$ and is assumed to be known and not-controllable. The reactive power from DERs, $q_{Dj}$, is assumed to be controllable and modeled as the decision variable for the resulting distribution-level optimal power flow problem. The apparent power rating of the smart inverter limits the reactive power dispatch ($q_{Dj}$) from the DER. {\color{black} Let the rating of the DER connected at node $j \in \N_D$ be $S_{DRj}$. Then the bounds on $q_{Dj}$ $\forall$ $j \in N_D$ are expressed using (2)}. 

\vspace{-0.3cm}
 \begin{small}
\begin{eqnarray}
\IEEEyesnumber\label{DG_lim} \IEEEyessubnumber*
-\sqrt{S_{DRj}^2-p_{Dj}^2} \leq q_{Dj} \leq \sqrt{S_{DRj}^2-p_{Dj}^2}
\end{eqnarray}
 \end{small}
 \vspace{-0.5cm}

\subsection{Centralized Non-linear and Relaxed OPF Problem}
A centralized OPF (C-OPF) problem is defined to optimize the network for some cost function characterized by a network-level problem objective, the power flow constraints in  \eqref{eqModel_nonlin1}-\eqref{eqModel_nonlin4}, and the operating constraints on the power flow variables. In this paper, we formulate the problem objective of minimizing network power losses. The aim is to reduce the network losses by controlling the reactive power output from DERs ($q_{Dj}$).

{\color{black}Let $X$ define the problem variables (including power flow and decision variables), where  $X=[P_{ij}$, $Q_{ij}$, $l_{ij}$, $v_j$, $q_{Dj}]^T$; $\forall j\in \N$, and $\forall (i,j)\in \E$. Let $F(X)$ denote the objective function representing the total power loss in the given distribution system. Note that $F(X)$ is a function of both the power flow variables and decision variables. Then, the centralized OPF problem is defined as the following in (P1).} 
 
\vspace{-0.4cm} 
\begin{small}
\begin{IEEEeqnarray}{C C}
\IEEEyesnumber\label{nonlin_OPF} \IEEEyessubnumber*
\text{(P1)}\hspace{0.4cm} \min \hspace{0.2cm} {\color{black}F(X)} = \sum_{(i,j)\in \E} l_{ij}r_{ij} \hspace{0.6cm}\\
\text{s.t.}  \hspace{0.2cm} \text{\eqref{eqModel_nonlin1} - \eqref{eqModel_nonlin4}, and \eqref{DG_lim}}\\
\underline{V}^2 \leq v_i \leq \overline{V}^2 \hspace{0.9cm} \forall i\in \N \label{nonlin_OPF1}\\
l_{ij} \leq \left(I^{rated}_{ij}\right)^2  \hspace{0.5cm} \forall (i,j) \in \E \label{nonlin_OPF2}
\end{IEEEeqnarray}
\end{small}
\vspace{-0.4cm}

\noindent where, $\overline{V} = 1.05$ and $\underline{V} = 0.90$ are the limits on bus voltages, and $(I^{rated}_{ij})^2$ is the thermal limit for the branch $(i,j)$. 

The NLP problem in (P1) can be reformulated as a convex optimization problem upon relaxing the quadratic equality constraints (1d), \eqref{eqModel_nonlin4}, as a convex inequality \cite{molzahn2019survey}. This relaxation leads to a Second-Order Cone Programming (SOCP) model, as detailed in (P2) using (4a)-(4c).  

\vspace{-0.4cm} 
\begin{small}
\begin{IEEEeqnarray}{C C}
\IEEEyesnumber\label{SOCP_OPF} \IEEEyessubnumber*
\text{(P2)}\hspace{0.4cm} \min \hspace{0.2cm} {\color{black}F(X)} = \sum_{(i,j)\in \E} l_{ij}r_{ij} \\
\text{s.t.}  \hspace{0.2cm} \text{\eqref{eqModel_nonlin1},\eqref{eqModel_nonlin2}, \eqref{eqModel_nonlin3}, \eqref{DG_lim}}, \text{\eqref{nonlin_OPF1}, and \eqref{nonlin_OPF2}}\\
v_il_{ij} \geq P_{ij}^2+Q_{ij}^2 \hspace{0.6cm} \forall (i,j)\in \E \label{SOCP_OPF1}
\end{IEEEeqnarray}
\end{small}
\vspace{-0.4cm}


{\color{black}\noindent \textbf{\noindent \textit{Remark 1:}} Note that for a balanced radial distribution system, the aforementioned SOCP relaxation is exact under specific sets of conditions \cite{farivar2013branch}. Also, whenever the relaxed mode is exact, \eqref{SOCP_OPF1} is active for the optimal solutions of the relaxed problem (P2). That is, the obtained solution using the relaxed model is also the optimal and power flow feasible for the original NLP. However, there may be cases when the solutions obtained from the relaxed SOCP model are not power flow feasible. These conditions are extensively studies in recent literature \cite{li2016non, Jha2019Exact}. In those cases, the NLP in (P1) needs to be solved to obtain the optimal and feasible power flow solution. For this reason, in the proposed D-OPF method, we solve the NLP model in (P1). However, we have introduced the SOCP model (P2) here to obtain the solution for the centralized OPF problem for the large test feeder used in simulations. This is because, for the large feeder, the NLP model did not converge. Thus, the solutions from the SOCP model were used for the comparison and validation purpose. Also, for the specific test case used in this paper, the SOCP solutions were found to be power flow feasible for the original NLP model and thus can be used for comparison of the solution quality.}

{\color{black}\noindent \textbf{\noindent \textit{Remark 2:}} It should be noted that the distribution system also includes legacy voltage control devices such as voltage regulators and capacitor banks. The legacy control devices, while not explicitly modeled here, can be easily incorporated in the proposed D-OPF model. However, upon incorporating such legacy devices, the optimization problem becomes Mixed-Integer Non-Linear Programming (MINLP) problem. Previously, we proposed a bi-level approach to manage the complexity of the MINLP problem resulting from the control of both legacy devices and smart inverters \cite{jha2019bi}. This algorithm can be adopted to solve the resulting MINLP problem in a computationally tractable manner.}

{\color{black} Furthermore, note that in the above formulation we have only considered reactive power from DGs ($q_{Dj}$) as controllable. However, the OPF formulation can be easily extended to the cases when both active ($p_{Dj}$) and reactive ($q_{Dj}$) power are controllable. In this case, the constraint on DER power rating defined in (2) simply transforms into a SOCP constraint.}

\begin{algorithm}[t] 
    \small
    \mbox{\emph{Algorithm 1: BCD Method}}\\
    \noindent\rule{8.2cm}{1pt}\\
    {Initialize the variable blocks $(x_1^0, x_2^0,...,x_s^0)$}\\
    And $k = 1$;\\
        \vspace{5pt}
    \While {stopping criterion is not met}    {
    \vspace{5pt}
    \For {$i = 1,2,...,s$}    {
    $x_i^{k}:= \underset{x_i \in \mathcal{S}_i}{\text{argmin}} \hspace{0.2cm} {\color{black}\Phi_i^k(x_i)}$
    }
    \vspace{3pt}
    $X^k = (x_1^k, x_2^k, ..., x_s^k)$\\
    $k \leftarrow k+1$
    \vspace{5pt}
    }
    {Return Global Minimizer: $X^* \leftarrow X^k$}\\
   \noindent\rule{8.2cm}{1pt}
\vspace{-1.0cm}
    \label{algori1}
\end{algorithm}

\vspace{-0.3cm}
\section{Distributed Optimal Power Flow (D-OPF)}
The loss minimization OPF problem defined in (P1) and (P2) is decomposable into connected areas. In this paper, it is assumed that the distribution system is composed of $N_A$ distributed areas. Let $\A_R$ be the set of all decomposed areas in a given distribution system, where $\A_R = \{ A_1, A_2, …, A_{N_A}\}$. Also, let each area, $A_m$, be defined as a directed graph  $A_m = \G(\N_m, \E_m)$, where $m = \{1,2,..., N_A\}$ that is controllable using a Local Controller (LC). The LCs solve a decomposed OPF problem to minimize the power losses for the specific area by controlling the local decision variables. Next, the proposed ENApp D-OPF framework is used to coordinate among the neighboring LCs and to facilitate network-level optimization. The proposed distributed approach is analogous to a primal-decomposition process that does an area-wise breakdown of the centralized problem and requires exchanging only primal variables \cite{langbort2004decomposition}. However, unlike the traditional primal-decomposition method, a new algorithm is proposed that leverages a reduced equivalent network representation for the power flow in radial distribution systems. The resulting sub-problems are solved using an approach similar to the \textit{Block Coordinate Descent (BCD)} method described in \cite{xu2013block}. 
The master problem requires solving only a set of non-linear equations of the form {\color{black}$f = 0$}. Here, we use the Fixed Point Iteration (FPI) method to solve the master problem. 

\vspace{-0.3cm}
\subsection{Block Coordinate Descent (BCD) Method}
In the BCD algorithm, the original problem variables are partitioned into several blocks. All but one block of the variables are fixed, and the problem is minimized over the free blocks. In this way, the problem is minimized for all the blocks. For example, the BCD method was used to solve the systems of equations \cite{ortega1970iterative,xu2013block}. 
The convergence properties of the BCD algorithm are reported in \cite{beck2013convergence,xu2017globally}. Generally, for a nonconvex problem, the BCD algorithm admits a sub-linear convergence. While, for a strongly convex optimization problem, the BCD algorithm is known to converge linearly. 

{\color{black}A brief discussion of the BCD method to solve optimization problems is described next. The objective is to solve the optimization problem detailed in \eqref{OP_com} that aims at minimizing the cost function $F(X)$, subject to the constraints defined by the constraint set $\mathcal{S}$. The BCD approach solves the optimization problem defined in (5) in a distributed way.  The approach to solving (5) using the BCD method is detailed in Algorithm 1  \cite{xu2013block}. Let variable $X$ be partitioned into $s$ blocks, where, $X = (x_1, x_2, ..., x_s)$. We define the modified cost function, $\Phi_i^k(x_i)$, for the $i^{th}$ block at $k^{th}$ iteration, in \eqref{BCD_f}. In the BCD approach, the decomposed cost function defined in (6) is minimized over one block while keeping other blocks constant. This process is repeated for other blocks until a prespecified stopping criterion is met. Thus, upon convergence, the solution for the original optimization problem defined in (5) is obtained in a distributed way.}
\begin{equation}
\small
\IEEEyesnumber\label{OP_com} \IEEEyessubnumber*
\min_{X} \hspace{0.2cm} F(X) \text{;}\hspace{0.3cm} \text{ s.t. }  \hspace{0.3cm}  \text{$ X \in \mathcal{S}$}
\end{equation}
\vspace{-0.3cm}
{\color{black}
\begin{equation}
\small
\IEEEyesnumber\label{BCD_f} \IEEEyessubnumber*
\Phi_i^k(x_i) = F(x_1^k, x_2^k, ..., x_{i-1}^k, x_i, x_{i+1}^{k-1},...,x_s^{k-1}) \text{ s.t. }  \hspace{0.1cm}  \text{$ x_i \in \mathcal{S}_i$}
\end{equation}
}


\begin{figure}[t]
     \vspace{-0.2cm}
    \centering
    \includegraphics[width=0.3\textwidth]{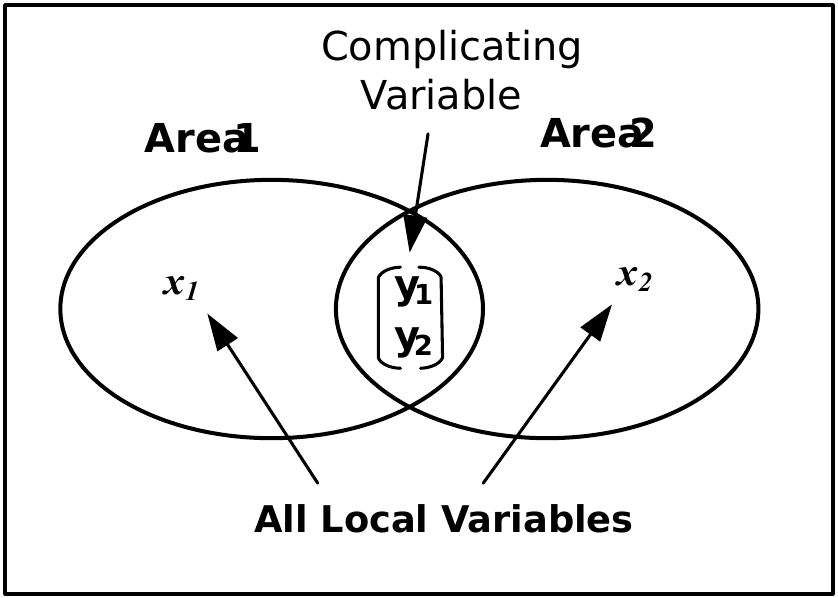}
 \vspace{-0.3cm}
    \caption{An example of a two-area system.}
    \label{two_area_proposed}
 \vspace{-0.5cm}
\end{figure}
\vspace{-0.6cm}
\subsection{Proposed ENApp Algorithm for D-OPF}
The developed D-OPF method leverages the radial topology of the power distribution network. Note that for a radial distribution system, there is a unique upstream area (UA) for any downstream area (DA). However, a UA (parent) can be connected to multiple DAs (child). We model a dummy bus for each UA-DA pairs; the dummy bus is shared by both the UA and DA regions. The dummy bus acts as a fixed voltage source for the DA while as a load (PQ) bus for the respective UA. This method is based on reduced network equivalents where the DA and UA are replaced by a fixed load and a fixed voltage source at each iteration, respectively. Each area solves its own OPF problem and then exchanges the boundary variables only with its immediate neighbors. Specifically, the DA shares complex power flow with the UA, while the UA shares bus voltages with the DA. The macro-iterations defining the exchange of variables among neighboring areas continue until both the UA and DA agree on the same value for the shared bus variables within a prespecified tolerance.

Note that the proposed D-OPF method is a modified version of the BCD algorithm for solving the sub-problems. However, in the proposed approach, the sub-problems for the distributed optimization modules are executed in parallel as opposed to the sequential execution adopted in the traditional BCD algorithm. Further, unlike the primal decomposition described in \cite{langbort2004decomposition}, the complicating/shared variables are split into two sets. One set of the shared variables are minimized in one of the subproblems while keeping the other set fixed. The process is then repeated for the other set of primal variables while keeping the first set fixed this time, similar to the BCD method. The iteration stops when the change in shared variables is within a prespecified tolerance.


\begin{algorithm}[t] 
    \small
    \mbox{\emph{Algorithm 2: ENApp D-OPF method for two Area System}}\\
    \noindent\rule{8.2cm}{1pt}
    
    {Initialize $\hat{y}^0_1 \in S_1$ and $\hat{y}^0_2 \in S_2$}
    
    {Initialize $\epsilon = 1, k = 1$}
    
    \While {$|\epsilon| > 0.001$}{
    Compute Parallely ($L_1, L_2$) --
    
    {$L_1:$ $X_1^k$ := $\underset{X_1 \in S_1}{\text{argmin}}\hspace{0.2 cm} \Phi_1(X_1, \hat{y}_2^{k-1})$}
    
    $L_2:\hspace{0.1cm}X_2^k :=  \underset{{\color{black}X_2 \in S_2}}{\text{argmin}} \hspace{0.2 cm} \Phi_2(X_2, \hat{y}_1^{k-1})$
    
    {Check residual}
    
    \vspace{3pt}
    $\mathcal{R}$ =
    $\left[
    \begin{array}{c}
    y_1^k - \hat{y_1}^{k-1}\\
    y_2^k - \hat{y_2}^{k-1}
    \end{array}
    \right]$
     \vspace{4pt}
     
    {FPI update:}
    
    \vspace{3pt}
    $\hat{y}_1^k = \psi_1(\hat{y}_2^{k-1},x_1^k)$
    
    \vspace{2pt}
    $\hat{y}_2^k =\psi_2(\hat{y}_1^{k-1},x_2^k)$

    \vspace{5pt}
    Increase iteration count $k$:
    
    $k \leftarrow k+1$
    
    \vspace{3pt}
    $\epsilon = max \hspace{3pt} |\mathcal{R}|$
        \vspace{10pt}
    
    }
    {Return Global Minimizer:  ${\{X^{\star}_{1}, X^{\star}_{2}\} = \{X^{k}_{1},X^{k}_{2}\}}$}
    
    \vspace{1pt}
   \noindent\rule{8.2cm}{1pt}
    \label{algori2}
        \vspace{-1.1cm}
\end{algorithm}

\subsubsection{D-OPF for Two-Area System}
First, we detail the proposed ENApp D-OPF algorithm using a two-area system. The two-area model is then extended to obtain the generalized model for the distribution system comprised of multiple areas. Suppose a network is composed of two areas: Area 1 (UA) and Area 2 (DA), each with their own local variables, $x_1$ and $x_2$ (see Fig. \ref{two_area_proposed}). 
Let $\textbf{Y} = [y_1, y_2]^T \in \mathcal{C}^2$ be the associated variable for the shared bus. Note that $\textbf{Y}$ is the complicating variable that couples the optimization problems for the two areas. Here, $y_1$ and $y_2$ are the bus voltage magnitude and the complex power flow through the bus shared between Area 1 and Area 2, respectively. 

Let the set of all local variables for Area 1 and Area 2 be $X_1$ and $X_2$, respectively where, $ X_1 = \{x_1, y_1\}{\hspace{0.2cm}} \text{\&{\hspace{0.2cm}}}X_2 = \{x_2, y_2\}$. Let $X$ be the set of all problem variables where, $X = X_1 \cup X_2$. The overall centralized optimization problem is defined using \eqref{eqProposed_obj1}, where, $S$ is the set of constraints, and $F$ is a decomposable cost function.
\begin{equation}\label{eqProposed_obj1}
\small
\min_{X \in S}{\hspace{0.2cm} F(X)}
\end{equation}
 The problem defined in (\ref{eqProposed_obj1}) can be written as \eqref{eqProposed_obj2}.
\begin{equation}\label{eqProposed_obj2}
\small
\min_{X_1 \in S_1, X_2 \in S_2}{\hspace{0.2cm}} \Phi_1(X_1,y_2)+\Phi_2(X_2,y_1)
\end{equation}
where, $S_1$ and $S_2$ are the set of constraints on local variables for Area 1 and Area 2, respectively. Here, $\Phi_1$, $\Phi_2$ are the cost functions for the respective local subproblems. 

Thus, the original problem defined in (\ref{eqProposed_obj2}) can be decomposed into the following two sub-problems.

\vspace{-0.3cm}
\begin{subequations}\label{eqProposed_subprb1}
\small
\begin{gather}
\min_{X_1 \in S_1}{\hspace{0.2cm}} \Phi_1(X_1,y_2) \label{sub1}\\
\min_{X_2 \in S_2}{\hspace{0.2cm} \Phi_2(X_2,y_1)} \label{sub2}
\end{gather}
\end{subequations}
\vspace{-0.3cm}

\begin{figure*}[t]
 \vspace{-0.5cm}
    \centering
    \includegraphics[width=0.95\textwidth]{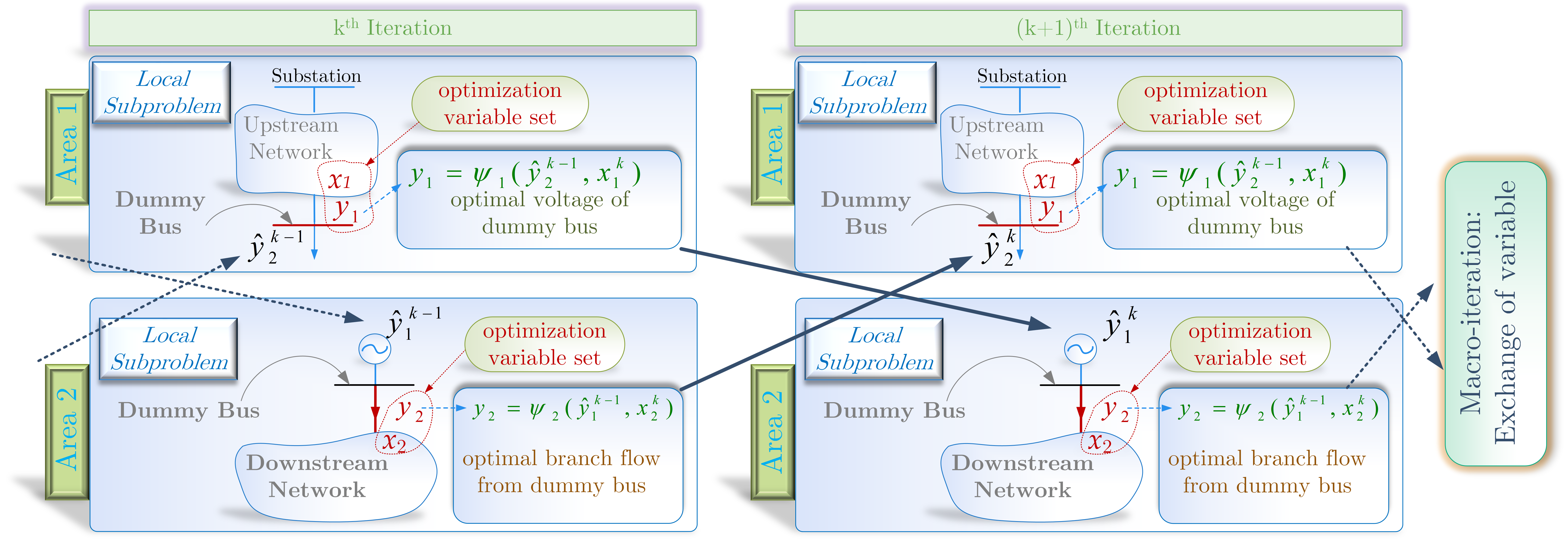}
    \centering
    \caption {Descriptive picture of Algorithm 2: ENApp Method.}
    \label{Algorithm Picture}
 \vspace{-0.5cm}
\end{figure*}

The subproblems defined in \eqref{eqProposed_subprb1} are solved as the following. The proposed distributed algorithm first solves the subproblems individually to obtain respective local and complicating variables. Here, $y_2$ and $y_1$ are kept fixed to solve subproblem \eqref{sub1} and sub-problem \eqref{sub2}, respectively for $X_1$ and $X_2$. The complicating variables, $y_1$ and $y_2$, are then exchanged among the two areas. The process is repeated until the master problem converges. The master problem is described using interface equations in \eqref{Master_prob} and solved using the FPI method. Note that the mathematical relation between $y_1$ and $y_2$ is formulated based on nonlinear power flow equations that are described as the interface equations in (10). The master problem converges when the change in complicating variable in successive iterations is within a pre-specified tolerance. The proposed algorithm for the two-area system is detailed in Algorithm 2 and can be easily extended to multiple areas as detailed later in this section. {\color{black} An overview figure representing the working of the proposed algorithm is shown in Fig. \ref{Algorithm Picture}. } 

\vspace{-0.3cm}
\begin{subequations}\label{Master_prob}
\small
\begin{gather}
y_1 - \psi_1(y_2,x_1) = 0\\
y_2 - \psi_2(y_1,x_2) = 0
\end{gather}
\end{subequations}
\vspace{-0.3cm}

\noindent \textbf{\noindent \textit{Remark 3 (Equivalent Network Approximation):}} The functions $\psi_1$ and $\psi_2$ in (10) project{\color{black}/map} the local variables ($x_1, x_2$) of higher dimensionality, to the shared variables ($y_1, y_2$) of lower dimensionality that represent network equivalents. {\color{black}The explicit form of these functions for the D-OPF problem is simply power flow equations. More explicitly, $\psi_1$ maps the boundary bus voltage with the power injections in the local area and the shared bus, and $\psi_2$ maps the boundary bus power flow with the power injections in the local area and voltage at the shared bus. Thus, the functions $\psi_1$ and $\psi_2$ are simply mapping the shared bus variables for the UA and DA areas using respective power flow equations for the local areas.} Exchanging boundary variables that correspond to network equivalents help achieve a faster convergence in fewer macro-iterations.

The network equivalents obtained in this work for UA and DAs are driven by the common practice assumptions for network equivalents in radial power distribution systems. Specifically, the aggregated load assumption to represent DA is a common practice in power systems analysis. It states that the downstream load of a bus can be approximated as a lumped load. This enables us to represent the voltage at the shared bus as $y_1 =\psi_1(y_2,x_1)$ where $y_2$ is the lumped load representing the network equivalent of the DA (Area 2) for the UA (Area 1). Similarly, as the power flow solutions are unique for radial distribution systems \cite{Baran_uniq}, $y_1$ represents the network equivalent for the UA (Area 1). Thus, the equivalent load at the shared bus, $y_2$, can be expressed as $y_2 =\psi_2(y_1,x_2)$, where $y_1$ is the bus voltage representing the network equivalent of the UA (Area 1) as seen from the DA (Area 2). {\color{black} Note that these network equivalents are not explicitly calculated. Rather, they are updated based on local optimization solutions at each macro-iterations.}

Thus, instead of all local variables, $x_1$ and $x_2$, we only need the shared variables, $y_1$ and $y_2$, during the macro-iterations to capture and share the information among distributed areas. Moreover, since the shared variables $y_1$ and $y_2$ represent electrical network equivalents for the respective UA and DA, it is able to provide more meaningful information to the respective regions during macro-iterations. This leads to a fewer number of variable exchanges to achieve convergence.



\begin{figure}[t]
 \vspace{-0.2cm}
    \centering
    \includegraphics[width=0.40\textwidth]{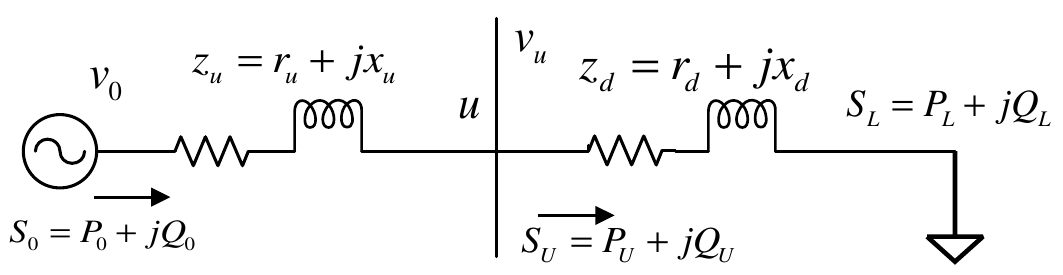}
    \caption{A simple radial distribution system.}
    \label{Simple_system}
 \vspace{-0.5cm}
\end{figure}

\subsubsection{D-OPF for a System with Multiple Areas}
For a system comprised of multiple areas, the OPF problem for the loss minimization objective is defined {\color{black} for each $A_m \in \A_R$ as shown in \eqref{Local_OPF}, where $m \in \{1,2,...,N_A\}$. Here, the problem variables for area $A_m$ are defined as $X_m =[P_{ij}$, $Q_{ij}$, $l_{ij}$, $v_j$, $q_{Dj}]^T$ $\forall j \in \N_m$ and $\forall (i,j) \in \E_m$. Recall that the reactive power dispatch from DERs, $q_{D_j}$, define the control/decision variables for the optimization problem.} 
The constraints for the OPF problem for area $A_m$ are defined in \eqref{Local_OPF_1}-\eqref{Local_OPF_socp}. Here, \eqref{Local_OPF_1}-\eqref{Local_OPF_socp} characterize the constraint set $\mathcal{S}_m$, for the subproblem defined for $A_m$.
{\color{black}Then, Algorithm 2 is implemented for each UA-DA pair. Note that we model a dummy bus between each connected area, which enables us to identify unique UA-DA boundaries. Thus Algorithm 2 can be extended for general radial distribution systems with multiple laterals.} The convergence is achieved when all the change in the shared variables for all area pairs is within a pre-specified tolerance.

\vspace{-0.3cm}
\begin{small}
\begin{IEEEeqnarray}{C C}
\IEEEyesnumber\label{Local_OPF} \IEEEyessubnumber*
\text{(P3)}\hspace{0.6 cm}\min \hspace{0.2cm}  \sum l_{ij}r_{ij} \\
\text{s.t.}\hspace{0.4 cm}P_{ij}-r_{ij}l_{ij}-p_{L_j}+p_{Dj}= \sum_{k:j \rightarrow k} P_{jk} \label{Local_OPF_1} \\
Q_{ij}-x_{ij}l_{ij}-q_{L_j}+q_{Dj}= \sum_{k:j \rightarrow k} Q_{jk}\\
v_j=v_i-2(r_{ij}P_{ij}+x_{ij}Q_{ij})+(r_{ij}^2+x_{ij}^2)l_{ij}\\
-\sqrt{S_{DRj}^2-p_{Dj}^2} \leq q_{Dj} \leq \sqrt{S_{DRj}^2-p_{Dj}^2}\\
\underline{V^2} \leq v_i \leq \overline{V^2} \\
{\color{black}l_{ij} \leq \left(I^{rated}_{ij}\right)^2} \\
v_il_{ij} = P_{ij}^2+Q_{ij}^2 \hspace{0.6 cm} \label{Local_OPF_socp}
\end{IEEEeqnarray}
\end{small}
\vspace{-0.3cm}


\vspace{-0.3cm}
\subsection{Convergence Analysis}
The network parameters play a vital role in the fast convergence of the proposed distributed algorithm. Here, for a simple system described in Fig. \ref{Simple_system}, the convergence at the boundary for solving the master problem (nonlinear power flow) is analyzed. In the system, load $S_L = P_L +jQ_L$ is connected to the node $u$ through a branch with $z_d = r_d+jx_d$ series impedance. A stiff voltage source at node $0$ with a squared voltage magnitude of $v_0$ is connected to the node $u$ through a series impedance of $z_u = r_u+jx_u$. The power flow from the voltage source to the system is $S_0 = P_0+jQ_0$. Also, the power flow from node $u$ is denoted by $S_u = P_u+jQ_u$. For the system (see Fig. \ref{Simple_system}), using the branch flow equations, the squared voltage magnitude of the boundary bus ($u$ node) can be written as \eqref{Con_1}.

 \vspace{-0.3cm}
\begin{small}
\begin{eqnarray}
\IEEEyesnumber\label{Con_1} \IEEEyessubnumber*
v_u = v_0 - 2(r_uP_0+x_uQ_0)+ z^2_u \frac{P_0^2+Q_0^2}{v_0}
\end{eqnarray}
\end{small}
 \vspace{-0.3cm}

\textbf{\textit{Assumption 1:}} To show convergence at the boundary bus, we assume that the $S_0$, $S_u$ and $S_L$ are in the order of $10^0 \sim 10^{-1}$; $|z_u|$ and $|z_d|$ are in the order of $10^{-2}$ or less. Note that these are valid assumptions for a practical power distribution system.

\textit{Lemma 1:}
$P_0^2+Q_0^2$ can be approximated by $P_u^2+Q_u^2$. Also, $P_0$ and $Q_0$ are in the same order as $P_u$ and $Q_u$, respectively.

\begin{proof}
By applying branch flow equation at the upstream bus, \eqref{lem1} can be written in terms of the line impedance and power injection at node $u$. The line loss is relatively small compared to the real and reactive power flow in the branch. By ignoring the line loss term, \eqref{lem1} can be written as \eqref{lem2}.

\vspace{-0.3cm}
\begin{small}
\begin{eqnarray}
\IEEEyesnumber\label{lem1} \IEEEyessubnumber*
\nonumber P_0^2+Q_0^2 &=& P_u^2+Q_u^2+ \Bigg\{\frac{P_0^2+Q_0^2}{v_0^2}[r_u^2+x_u^2]\Bigg\} \\&+&\Bigg\{\frac{2(P_0^2+Q_0^2)}{v_0}[r_uP_u+x_uQ_u]\Bigg\}\\
\IEEEyesnumber\label{lem2} \IEEEyessubnumber*
\frac{P_0^2+Q_0^2}{v_0} &=& \frac{P_u^2+Q_u^2}{v_0 - 2\{r_uP_u+x_uQ_u\}}
\end{eqnarray}
\end{small}
\vspace{-0.3cm}

Since $v_0 \approx 1$ and branch power losses are relatively smaller than the branch power flow for a well-designed power system \cite{farivar2013branch}, $v_0 >> 2\{r_uP_u+x_uQ_u\}$, and $P_0^2+Q_0^2 \approx P_u^2+Q_u^2$. 
\end{proof}

Again, for the downstream area, assuming that the branch power losses are relatively less compared to the branch power flow \cite{farivar2013branch}, we can express $P_u$ and $Q_u$ in terms of $P_L$ and $Q_L$ as \eqref{Con_2}.

\vspace{-0.3cm}
\begin{small}
\begin{eqnarray}\label{Con_2}
P_u = P_L + r_d \frac{P_L^2+Q_L^2}{v_u} \hspace{0.2cm} \text{and} \hspace{0.2cm}
Q_u = Q_L + x_d \frac{P_L^2+Q_L^2}{v_u}
\end{eqnarray}
\end{small}
\vspace{-0.3cm}

Using Lemma 1, \eqref{Con_1}, and \eqref{Con_2}, $v_u$ can be written as \eqref{B_F}. Here, at $k$ iteration, the voltage update step for the shared bus of UA uses the power injection at the same bus obtained from the DA at $({k-1})^{th}$ iteration. After  incorporating the iteration indices in \eqref{B_F}, we obtain \eqref{Con_3}.

\vspace{-0.3cm}
\begin{small}
\begin{eqnarray}
\IEEEyesnumber\label{B_F} \IEEEyessubnumber*
v_u = v_0 - 2(r_uP_u+x_uQ_u)+ z^2_u \frac{P_u^2 +Q_u^2}{v_0}
\end{eqnarray}
\begin{eqnarray}
\nonumber v_u^{k} &=& v_0 - 2(r_uP_u^{k-1}+x_uQ_u^{k-1})+ z^2_u \frac{(P_u^{k-1})^2+(Q_u^{k-1})^2}{v_0}\\
\nonumber &=& v_0 - 2\left\{r_uP_L + x_uQ_L + (r_ur_d+ x_ux_d) \frac{S_L^2}{v_u^{k-1}}\right\}\\
\nonumber && + \frac{z_u^2}{v_0}\left\{S_L^2+\frac{ z_d^2S_L^4}{(v_u^{k-1})^2} + \frac{2S_L^2}{v_u^{k-1}} \Big(r_dP_L +x_dQ_L \Big)\right\} \\
\nonumber &\approx& v_0 - 2(r_uP_L + x_uQ_L)+\frac{z_u^2S_L^2}{v_0} +\frac{2S_L^2}{v_u^{k-1}}\Big\{ \frac{z_u^2}{v_0}(r_dP_L +...\\
\IEEEyesnumber\label{Con_3} \IEEEyessubnumber*
&&x_dQ_L) -(r_ur_d+ x_ux_d) \Big\} \hspace{0.3 cm} \text{ (ignoring $z_u^2z_d^2$ term)  }
\end{eqnarray}
\end{small}
\vspace{-0.3cm}

Next, we define $C_1=v_0 - 2(r_uP_L + x_uQ_L)+\frac{z_u^2S_L^2}{v_0}$ and $C_2 = 2S_L^2\big\{\frac{z_u^2}{v_0}(r_dP_L +x_dQ_L)-(r_ur_d+ x_ux_d) \big\}$. Then the voltage at the shared bus at $k^{th}$ iteration, $v_u^k$, can be written as a function of the bus voltage at the previous iteration, $v_u^{k-1}$, \eqref{Con_4}. 

\vspace{-0.3cm}
\begin{small}
\begin{eqnarray}
\IEEEyesnumber\label{Con_4} \IEEEyessubnumber*
v_u^{k} = C_1 + \frac{C_2}{v_u^{k-1}}
\end{eqnarray}
\end{small}
\vspace{-0.3cm}

The speed of the convergence of the sequence $\{v_u^k\}$, defined by \eqref{Con_4}, depends on the $\gamma = \frac{C_2}{C_1}$. Here, for any power distribution system, $|\gamma|$ is bounded between $(0,1)$. The closer the $|\gamma|$ is to the zero, the faster the sequence converges. The convergence is guaranteed regardless of the value of $\gamma$. 

The ratio between the difference of two consecutive terms in the sequence $\{v_u^k\}$ can be expressed as \eqref{Con_sp1} and reduced to \eqref{Con_sp2} after some algebraic manipulations.

\vspace{-0.3cm}
\begin{small}
\begin{eqnarray}
\IEEEyesnumber\label{Con_sp1} \IEEEyessubnumber*
\frac{\left|{v_u^{k+2}-v_u^{k+1}}\right|}{\left| v_u^{k+1}-v_u^{k}\right|} = \frac{\left| C_1+ \frac{C_2}{v_u^{k+1}}-v_u^{k+1}\right|}{\left| v_u^{k+1}-\frac{C_2}{v_u^{k+1}-C_1}\right| }=\frac{\left|v_u^{k+1}-C_1\right|}{\left| v_u^{k+1}\right| }
\end{eqnarray}

\vspace{-0.3cm}
$$\frac{\left|{v_u^{k+2}-v_u^{k+1}}\right|}{\left| v_u^{k+1}-v_u^{k}\right|} =\frac{\left| C_1+ \frac{C_2}{v_u^{k}} -C_1\right|}{\left| C_1+ \frac{C_2}{v_u^{k}}\right| }=\frac{\left| \frac{C_2}{C_1}\right|}{\left|v_u^{k}+ \frac{C_2}{C_1}\right| }$$
\vspace{-0.3cm}
\begin{eqnarray}
\IEEEyesnumber\label{Con_sp2} \IEEEyessubnumber*
\hspace{0.8cm}=\frac{\left| \gamma\right|}{\left|v_u^{k}+ \gamma\right| } < 1
\end{eqnarray}
\end{small}

\vspace{-0.3cm}
From \eqref{Con_sp2}, it is clear that the convergence is guaranteed for \eqref{Con_4} {\color{black}since $v_u^k \approx 1$ and $|\gamma|$ is bounded between $(0,1)$. The convergence is linear as there exists some $\delta \in (0,1)$ for which $\frac{\left| \gamma\right|}{\left|v_u^{k}+ \gamma\right| } \le \delta$. Also, for any practical power distribution system system, the line parameters such as $r_d$, $x_d$ etc. are much smaller (in the order of $10^{-1}$ pu or smaller) than the allowable voltage magnitude ($\sim 1$ pu). Thus, $|\gamma|$ is at-most 0.1 for power distribution network. The rate of convergence depends on $|\gamma|$; the lower the $|\gamma|$, the higher the rate of convergence.} This property results in a faster convergence of the proposed algorithm at the boundary bus requiring a significantly fewer number of  macro-iterations or communication-rounds among distributed agents.

Fig. \ref{Simple_system} shows an equivalent representation of the two-area problem. The upstream and downstream branch can be thought as an equivalent circuit model of the UA and DA, respectively. Since in the ENApp D-OPF model, convergence is searched at the boundary bus of the physically neighboring areas, this analysis is valid for a multi-area system as well.

\vspace{-0.3cm}
\section{Results and Discussions}
The proposed distributed OPF approach is thoroughly validated using IEEE 123-bus and IEEE-8500 node test systems. All experiments were simulated in Matlab 2018b on a machine with 16GB memory and Core i7-8550U CPU @1.80GHz. The NLP problems in both C-OPF and D-OPF models are solved using a commercial NLP solver, \textit{Artelys Knitro} \cite{nocedal2006knitro}.  

\vspace{-0.3cm}
\subsection{Simulated System}
\begin{figure}[t]
  \centering
  \includegraphics[width=0.42\textwidth]{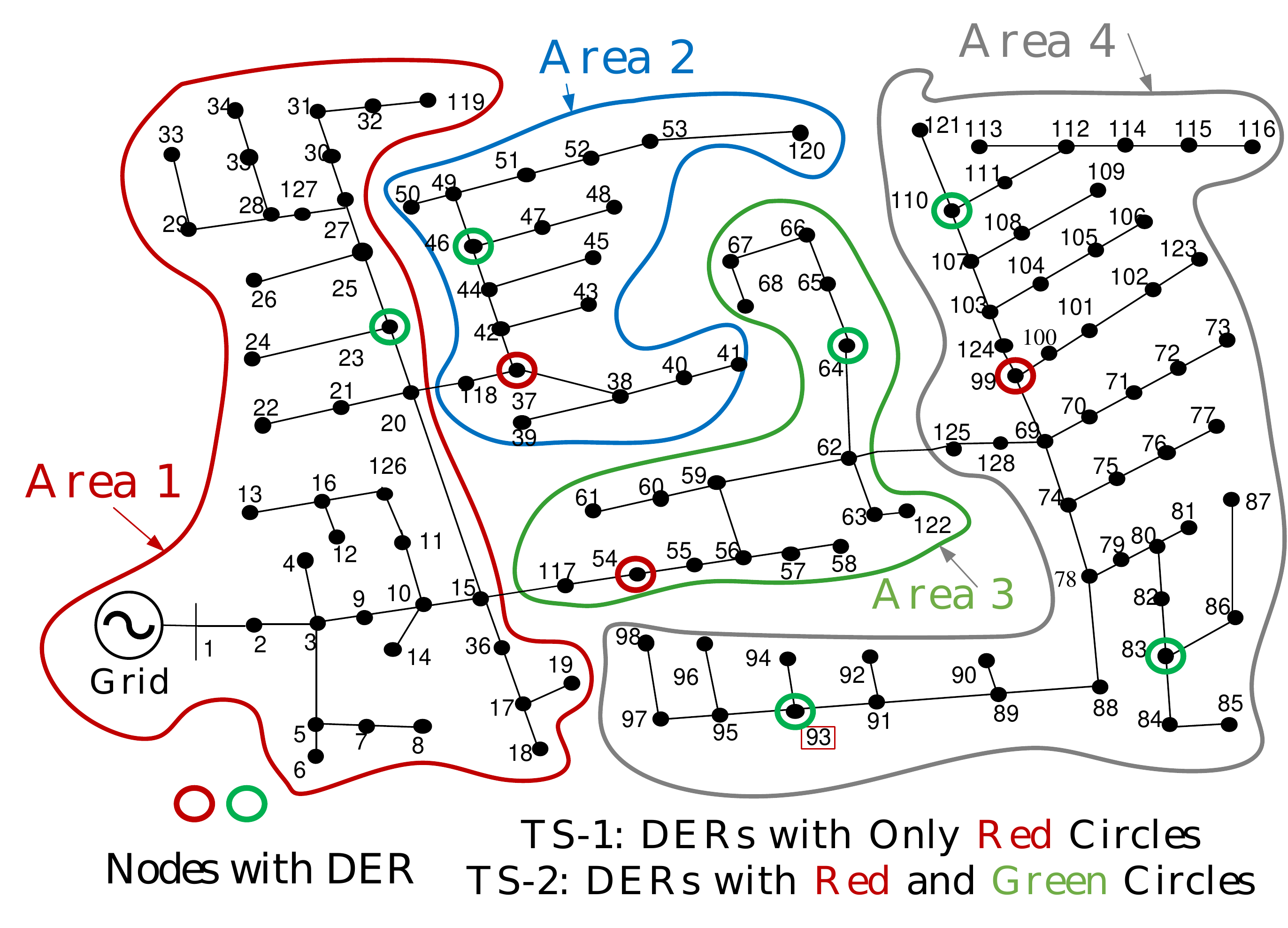}
     \vspace{-0.2cm}
 \caption{TS-1 and TS-2 Test System}
\label{IEEE_123Test_area_dec}
\vspace{-0.2cm}
\end{figure}

\begin{figure}[t]
  \centering
  \includegraphics[width=0.32\textwidth]{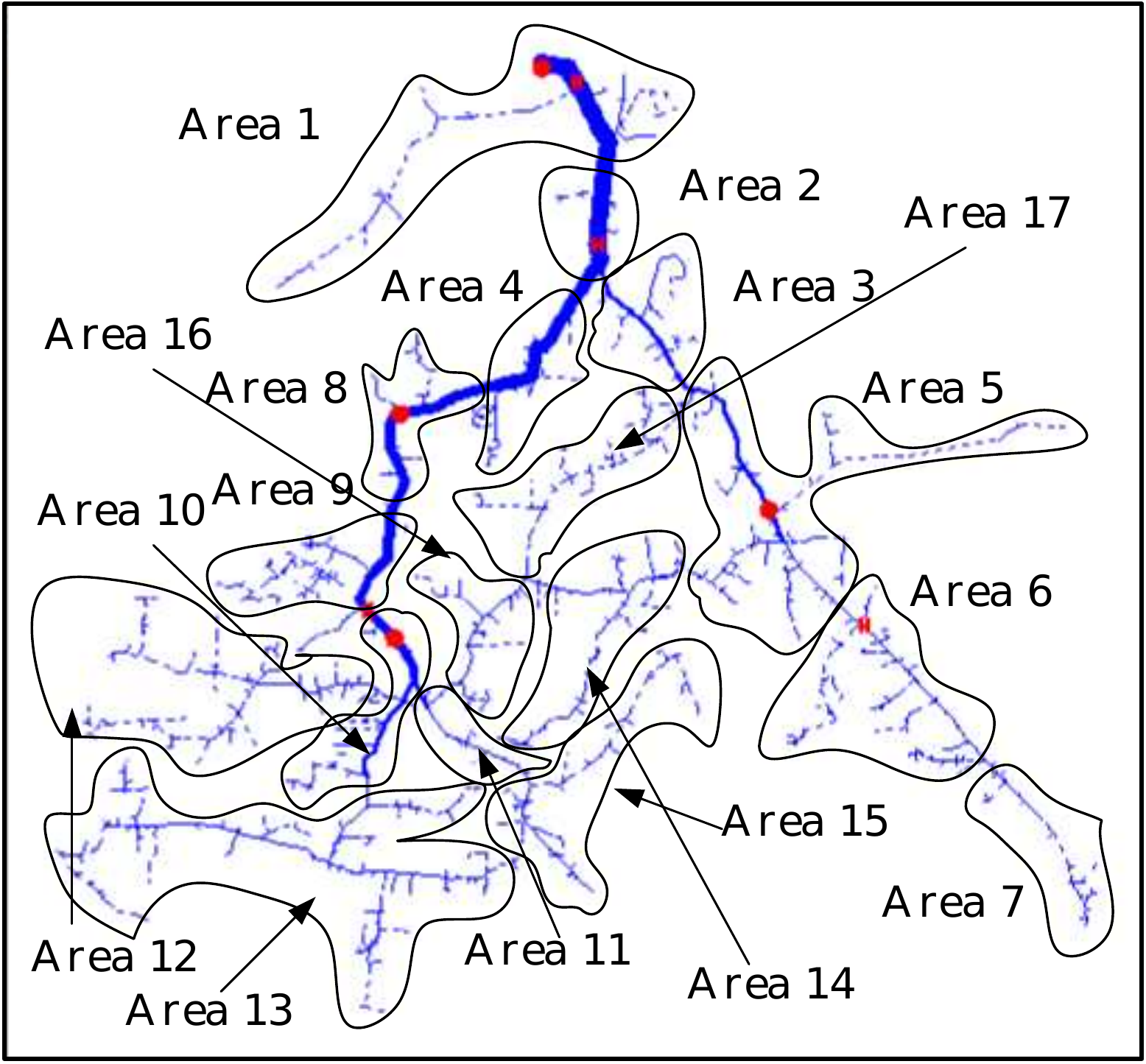}
 \caption{TS-3 and TS-4 Test System}
\label{IEEE_8500Test_area_dec}
\vspace{-0.5cm}
\end{figure}

The simulations are conducted using the following four test systems: (1) IEEE 123 bus system with 3 DERs (TS-1), (2) IEEE 123 bus with 9 DERs (TS-2), (3) IEEE 8500 node test system with 113 DERs - about $10\%$ ($|\N_D|/|\N_L| = 0.1$) DER penetration  (TS-3),and (4) IEEE 8500 node test system with 544 DERs - about $50\%$ ($|\N_D|/|\N_L| = 0.5$) DER penetration (TS-4).  {\color{black} For IEEE 123-bus system, the kVA rating of the DERs is 60 kVA. For IEEE 8500 node test system, the DER sizes are chosen randomly with a kVA rating ranging from 1.3 to 5.8 kVA. The simulation parameters associated with the respective test systems are shown in Table \ref{System_variables}.} 
Here, TS-3 and TS-4 are used to test the scalability of the proposed ENApp D-OPF method for larger distribution systems. We have simulated a single-phase equivalent circuit models for both IEEE 123-bus and IEEE 8500-bus test systems.
It is assumed that the test systems are already decomposed and comprised of multiple areas. The decomposition of both test systems and the connections between areas are shown in Fig. \ref{IEEE_123Test_area_dec} and \ref{IEEE_8500Test_area_dec}. 
TS-1 and TS-2 are comprised of four areas (Fig. \ref{IEEE_123Test_area_dec}) and TS-3 and TS-4 are comprised of 17 areas (see Fig. \ref{IEEE_8500Test_area_dec}).

The proposed distributed OPF algorithm is validated for optimality upon comparing against an equivalent centralized OPF model. We also compare the compute time requirements for the proposed distributed OPF and the centralized OPF models. Centralized OPF is formulated using the nonlinear branch flow model as detailed in (P1) (see Section II.B). 

\begin{table}[t]
\renewcommand{\arraystretch}{1.4}
\centering
\caption{{\color{black} Summary for Simulated Systems}}
\resizebox{0.49\textwidth}{!}
{\begin{tabular}{ |c|c|c|c|c|c|}
\hline
Test System & Area &Node &Total Variables & Constraints &DERs\\
\hline
TS-1 & 4 & 128 & 512& 898 &3\\
\hline
TS-2 & 4 & 128 & 518 & 910&9\\
\hline
TS-3 & 17 & 2522 & 10198 &17876 &113\\
\hline
TS-4 & 17 & 2522 & 10692 &18738 &544\\
\hline
\end{tabular}}
\label{System_variables}
\end{table}

\vspace{-0.3cm}
\subsection{IEEE 123-bus Test System}
In this section, we present the results for the IEEE 123-bus system using TS-1 (with 3 DERs) and TS-2 (with 9 DERs). The OPF problem is solved for the loss minimization problem. The results obtained from the proposed D-OPF method are compared with the C-OPF problem defined in (P1). Here, we use a small test system to demonstrate the convergence properties of the proposed D-OPF algorithm. 

For the 123-bus test system, the value of the objective function for the loss minimization optimization problem and time required for C-OPF and ENApp D-OPF to converge are compared in Table \ref{non_lin_comp_loss_123bus}. The table also shows the number of macro-iterations it takes for the proposed algorithm to converge. For the D-OPF approach, the compute time is the total time it takes for the D-OPF to converge given each area is computing the local optimization solutions in parallel. Note that the proposed approach converges to the same optimal solution as that obtained using the C-OPF algorithm (see Table \ref{non_lin_comp_loss_123bus}). Thus the proposed approach is able to achieve the optimal solution. Also, C-OPF takes 4.3 seconds to converge. D-OPF takes 4 macro-iterations and 3 seconds to converge. For the small system, while the computational advantages are not apparent, the major advantage is in the number of macro-iterations that ENApp D-OPF takes to converge to the optimal solution; it takes 4 macro-iterations to solve the non-linear OPF problem for TS-1 and TS-2. Notice that the existing literature reports that the state-of-the-art distributed optimization methods take hundreds of macro-iterations to converge for the IEEE 123-bus test system \cite{zheng2015fully,peng2016distributed,cavraro2017local}. {\color{black} We further provide additional comparison case studies to better justify this statement (see Section IV.D).}

\begin{table}[t]
\centering
\caption{C-OPF vs. ENApp D-OPF for IEEE 123-bus Test System}
{\begin{tabular}{ |c|c|c|c|c|c|}
\hline
\multirow{2}{*}{Test System}&  \multicolumn{2}{c|}{Total Loss (kW)} & \multicolumn{2}{c|}{Time (sec)}& \multicolumn{1}{c|}{\#Iterations}\\\cline{2-6}
 &{C-OPF}&{D-OPF}&{C-OPF}&{D-OPF}&{D-OPF}\\
\hline
{TS-1}& {21.7}&{21.9} & {4.3}& {3} & 4\\
\hline
{TS-2}&{12.6} & {12.6} & {7.94} & {6.47} & 4\\
\hline
\end{tabular}}
\label{non_lin_comp_loss_123bus}
\end{table}

\begin{figure}[!t]
\centering
\includegraphics[width=0.23\textwidth]{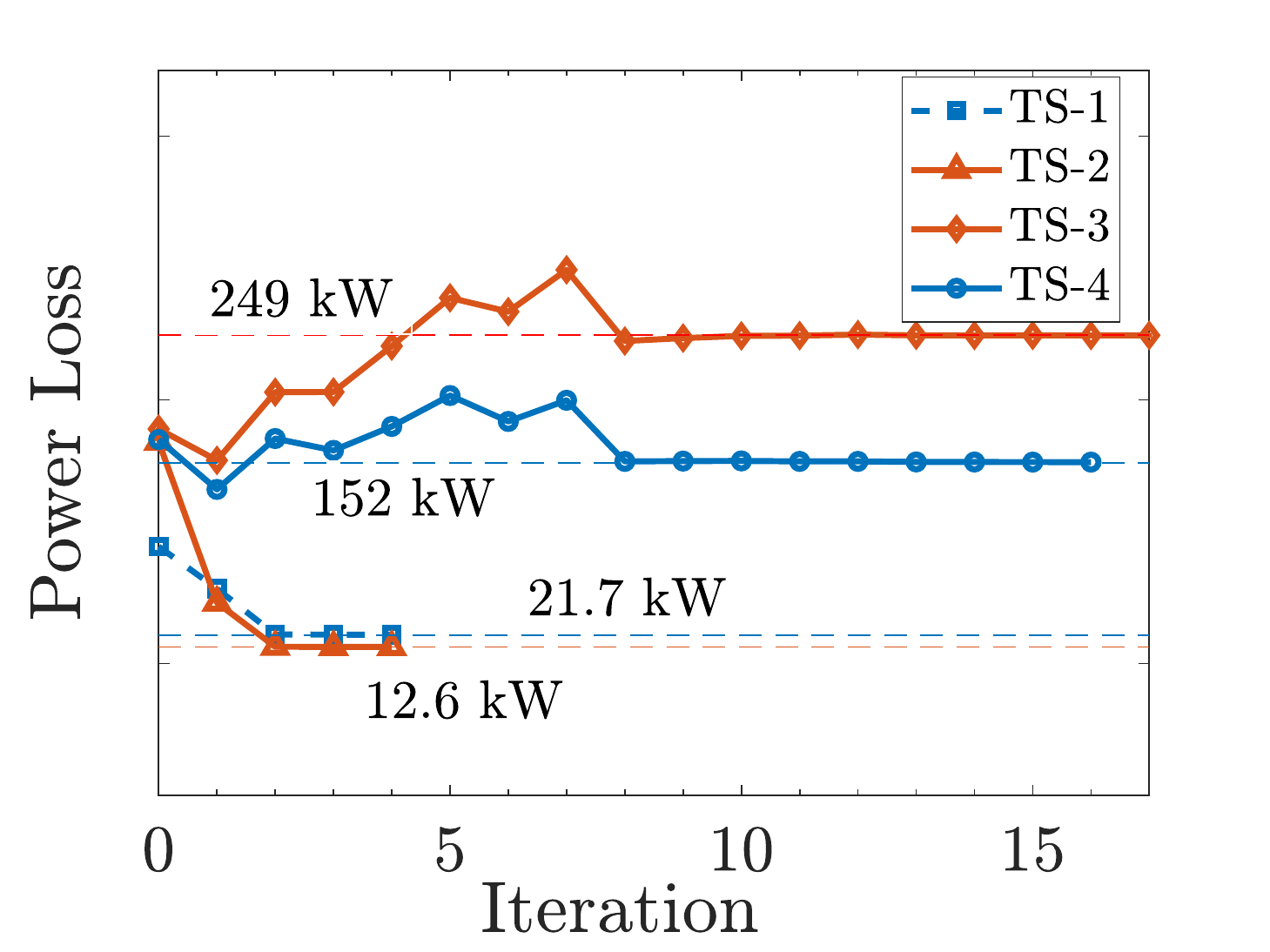}
\includegraphics[width=0.23\textwidth]{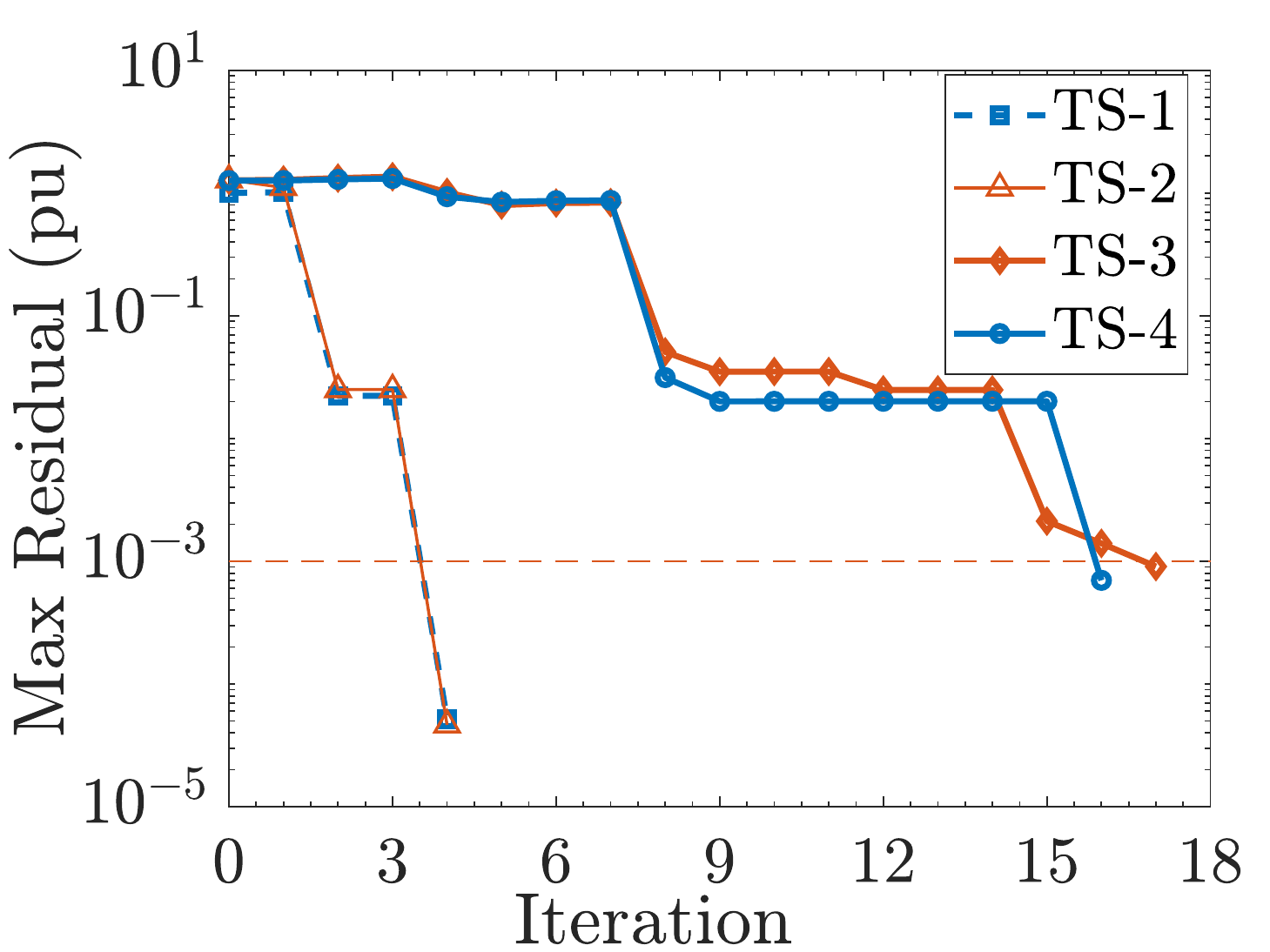}
\vspace{-0.2cm}
\caption{\centering{ENApp D-OPF Convergence Properties. At each macro-iteration (a) Objective function; and (b) Maximum residual of boundary variables.}}
\label{results_testcases}
\vspace{-0.5cm}
\end{figure}

\begin{figure}[!t]
\centering
\includegraphics[width=0.25\textwidth]{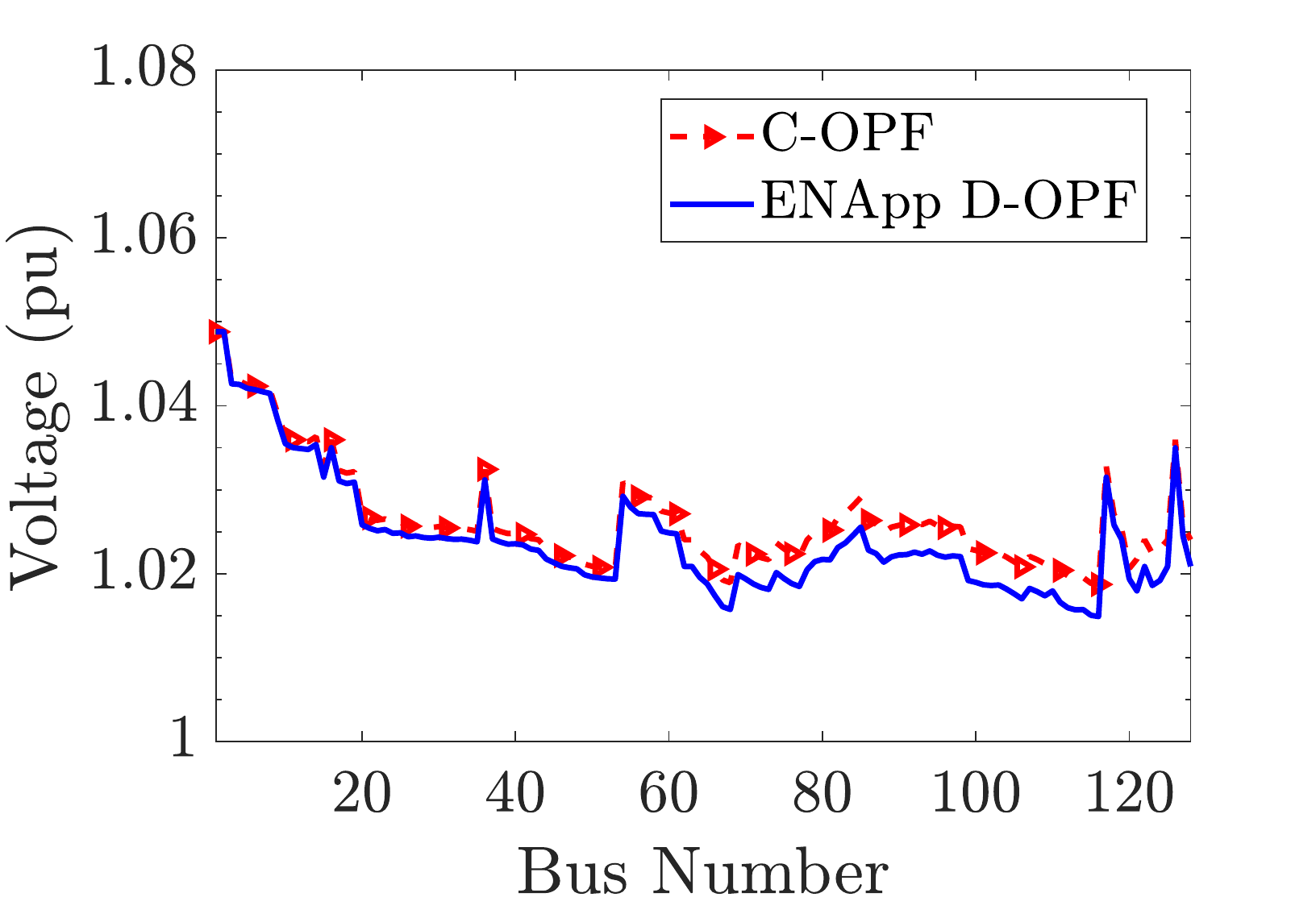}%
\includegraphics[width=0.25\textwidth]{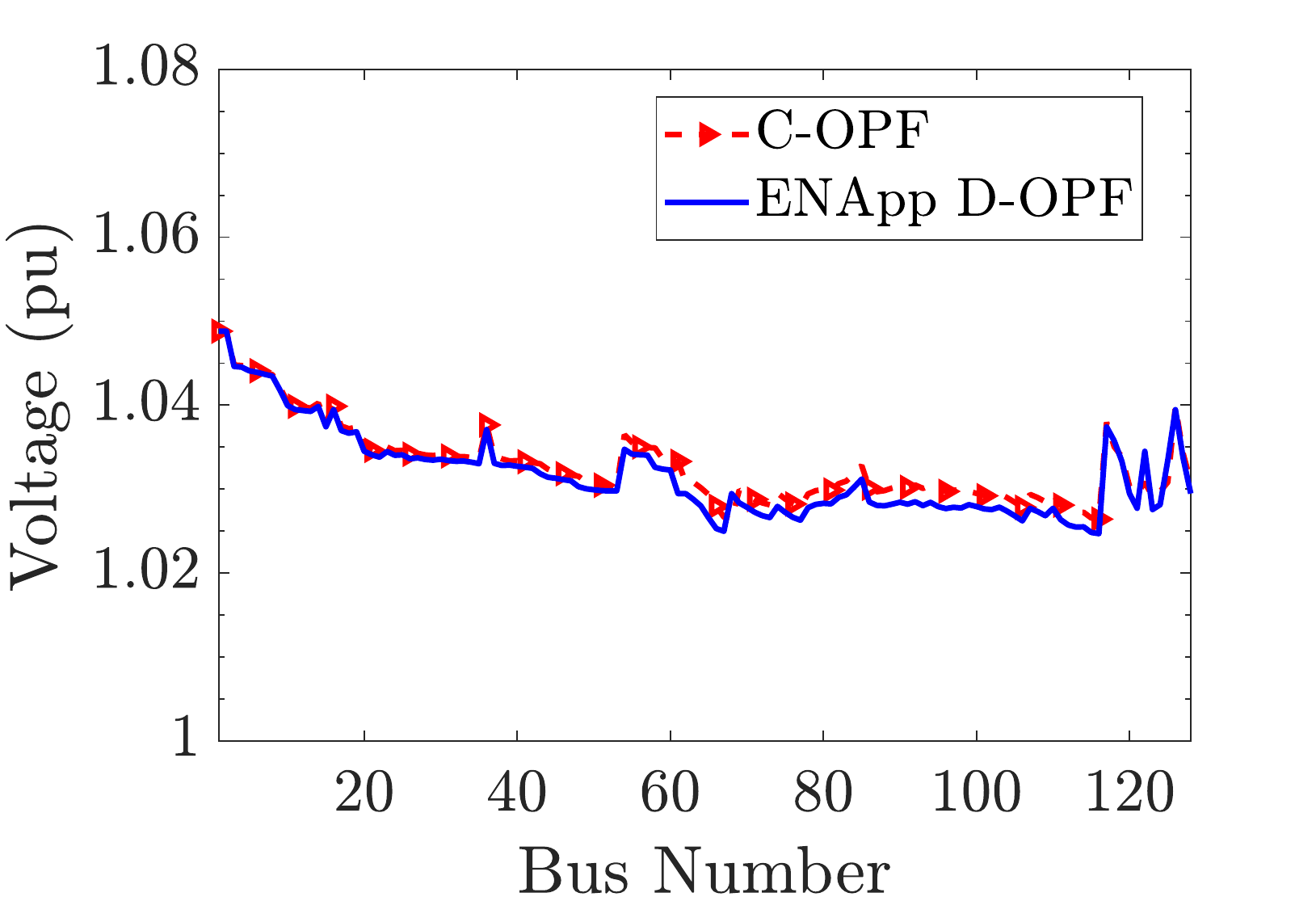}%
\vspace{-0.2cm}
\caption{\centering{Comparison of Node Voltages for proposed ENApp D-OPF and C-OPF Method, for (a) TS-1, (b) TS-2}}
\label{Comparing_voltages_123bus}
\vspace{-0.4cm}
\end{figure}

Next, the value of the objective function and the maximum residuals for boundary variables for each macro-iteration of the D-OPF are shown for each test system in Fig. \ref{results_testcases}. It can be observed that the D-OPF takes less than 4 iterations to converges to the optimal solutions for the IEE 123-bus test system, see Fig. 6(a). The residual for the boundary variables also rapidly decreases, see Fig. 6(b). The nodal voltages obtained using C-OPF and ENApp D-OPF method are also compared and shown in Fig. \ref{Comparing_voltages_123bus}. It can be observed that the ENApp D-OPF converges to the same optimal solution as obtained by solving the centralized OPF problem. Also, the difference in the nodal voltages obtained using C-OPF and the proposed D-OPF is $<0.1\%$.

\vspace{-0.3cm}
\subsection{IEEE 8500-node System}
We demonstrate the scalability of the proposed approach using the single-phase equivalent of IEEE 8500-node system with 2522 buses. We simulate two test scenarios using the IEEE 8500-node test system: TS-3 with 113 DERs and TS-4 with 544 DERs. Note that this test case comprises of a large system for OPF studies with a couple of thousand decision variables. Typically, the existing distributed OPF algorithms have not been tested for a system of this size and with these many DERs. Further, note that while the non-linear C-OPF (P1) problem is easily solved for the small test system such as IEEE 123 bus system, the commercial NLP solvers such as \textit{Artelys Knitro} \cite{nocedal2006knitro} fail to converge for the selected large test system. In fact, one of the major disadvantages of the large-scale centralized optimal power flow (C-OPF) problems is the long compute time or failure to achieve convergence. Thus, for validation, we compute optimal results using the centralized SOCP relaxation of the NLP problem as detailed in (P2), {\color{black}denoted by C-OPF$^*$}. {\color{black}Note that while, in general, SOCP solutions may not be power flow feasible \cite{li2016non}, they are found to exact for the loss minimization problems under certain conditions \cite{farivar2013branch2}. In this case, we explicitly validate that the SOCP solution is also power flow feasible. Thus, for this case, the SOCP solution provides a suitable optimal solution for comparison.} We use \textit{cvx} package to solve the relaxed SOCP problem \cite{grant2014cvx}.


\begin{table}[t]
\centering
\caption{C-OPF vs. ENApp D-OPF for IEEE 8500-node Test System}
{\begin{tabular}{ |c|c|c|c|c|c|}
\hline
\multirow{2}{*}{Test System}&  \multicolumn{2}{c|}{Total Loss (kW)} & \multicolumn{2}{c|}{Time (sec)}& \multicolumn{1}{c|}{\#Iterations}\\\cline{2-6}
 &{C-OPF{\color{black}$^*$}}&{D-OPF}&{C-OPF}&{D-OPF}&{D-OPF}\\
\hline
{TS-3} &{249}&{249.1} & {NA}& {180} & 17\\
\hline
{TS-4} &{152} & {152.8} & {NA} & {700} & 17\\
\hline
\end{tabular}}
\label{non_lin_comp_loss}
\end{table}

The results are shown in Table \ref{non_lin_comp_loss}. It can be observed that both C-OPF and D-OPF converge to the same optimal solution; the error for the value of the minimized objective function for the ENApp D-OPF method is less than 1\%. The compute time is not reported for the C-OPF problem as the NLP model did not converge for the 8500-node test system. Therefore, we report the optimal values obtained using the relaxed SOCP model for the C-OPF problem for comparison. On the contrary, the proposed D-OPF algorithm is able to solve the nonlinear programming model for the OPF detailed in (P3) and is able to achieve convergence relatively fast for the large system. Furthermore, the major advantage lies in the number of macro-iterations it takes to converge for the 8500-node test system. Even for the large test system, the proposed D-OPF achieves convergence using a relatively smaller number of macro-iterations, here, 17 macro-iterations. Thus, unlike other approaches in the literature, the number of macro-iterations increases sub-linearly with the number of decomposed areas in the test system. To the knowledge of the authors, no other distributed optimization methods reported in the literature can solve the non-linear OPF problem and reach the optimum solution in fewer macro-iterations for large-scale distribution systems (TS-3 or TS-4). The existing methods take hundreds of rounds of communication at each step of the optimization problem for large-scale distribution networks \cite{zheng2015fully,peng2016distributed}. Thus, this test case demonstrates the scalability of the proposed distributed optimization method for non-linear OPF problems.

\begin{figure}[t]
\centering
\includegraphics[width=0.25\textwidth]{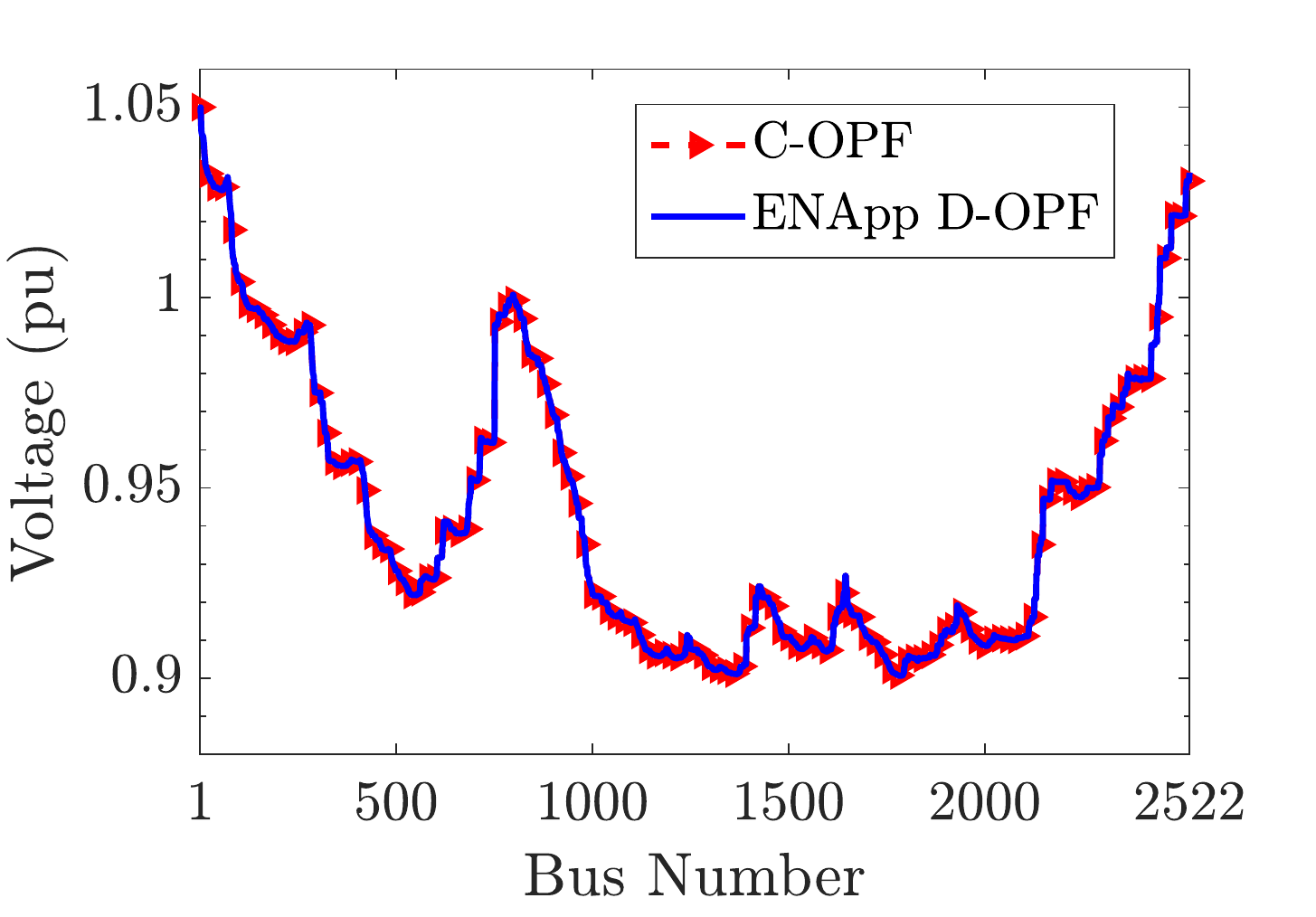}%
\includegraphics[width=0.25\textwidth]{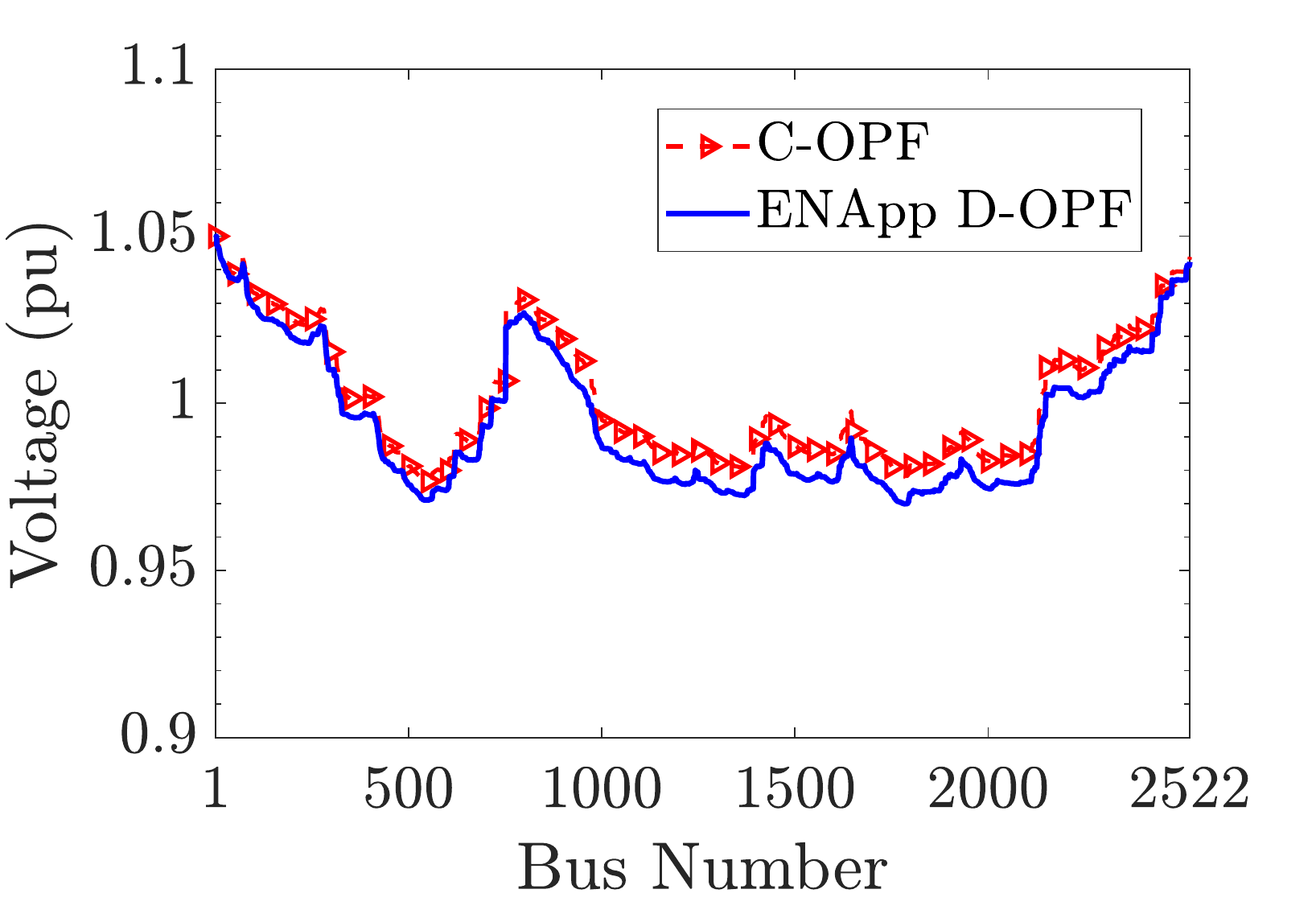}%
\caption{\centering{Comparison of Node Voltages for proposed ENApp D-OPF and C-OPF Method, for (a) TS-3, (b) TS-4}}
\label{Comparing_voltages}
\vspace{-0.4cm}
\end{figure}

The value of the objective function and the maximum residuals for boundary variables with respect to the number of macro-iterations for the simulated test systems are shown in Fig. \ref{results_testcases}. Note that the computation starts with a flat start and still converges within a reasonable number of iterations. Further, the maximum residual vector rapidly decreases with the iteration count. Next, the nodal voltages obtained using C-OPF and ENApp D-OPF method are compared in Fig. \ref{Comparing_voltages}. From the figure, it is clear that the proposed ENApp D-OPF can solve the original optimization problem regardless of the network size, and the error in voltage magnitude is very low ($<0.1\%$). 

For completeness, we validate the feasibility of the SOCP solutions for the simulation test cases TS-3 and TS-4. Note that the SOCP relaxation is exact if the solution for the relaxed problem satisfies the equality conditions for the relaxed equality constraint in (P2). That is, $e_{ij} = v_il_{ij} - (P_{ij}^2+Q_{ij}^2) \approx 0,$  ${\forall {ij} \in \E}$. It is observed that $e_{ij}$ is in the order of $10^{-4}$ upon solving SOCP relaxed C-OPF problem for the IEEE 8500-node test system (for both test systems, TS-3 \& TS-4). Thus, the SOCP solution is exact for the NLP OPF problem and can be used to benchmark the optimum solutions obtained from the proposed D-OPF algorithm. 

\begin{figure}[t]
\centering
\includegraphics[width=0.35\textwidth]{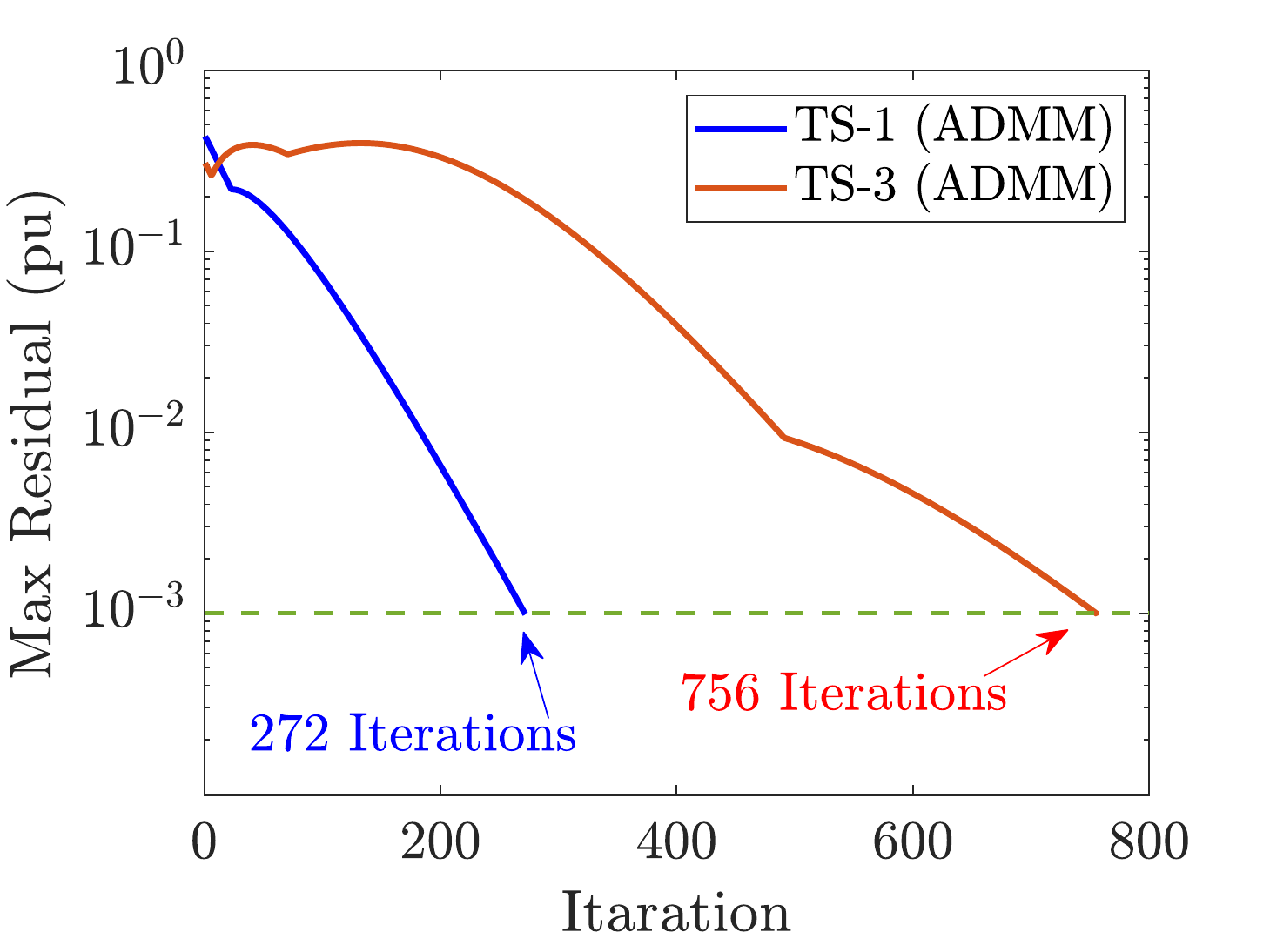}%
\vspace{-0.2cm}
\caption{\centering{Convergence Properties for ADMM method \cite{zheng2015fully} for TS-1 and TS-3}}
\label{ADMM_Rs}
\end{figure}

\begin{table}[!t]
\renewcommand{\arraystretch}{1.4}
\centering
\caption{{\color{black}Comparison of ENApp D-OPF and ADMM Based Method}}
\resizebox{0.49\textwidth}{!}
{\begin{tabular}{ |c|c|c|c|c|c|c|}
\hline
\multirow{2}{*}{System}&  \multicolumn{2}{c|}{Total Loss (kW)} & \multicolumn{2}{c|}{Time (sec)}& \multicolumn{2}{c|}{\#Iterations}\\\cline{2-7}
 &{ENApp}&{ADMM}&{ENApp}&{ADMM}&{ENApp}&{ADMM}\\
\hline
{TS-1}&{21.9}&{21.8}&{3}&{145}&{4}&{272}\\
\hline
{TS-3}&{249.1}&{249.1}&{180}&{6120}&{17}&{756}\\
\hline
\end{tabular}}
\label{ADMM_TS13}
\end{table}
{\color {black}
\subsection{Comparison with ADMM}
As stated before, the state-of-the-art distributed optimization approaches such as, ADMM based methods require a large number of communication rounds (macro-iterations) among distributed agents, which is in the order of $~10^2$ for mid-size networks. 
Next, we compare the proposed ENApp D-OPF method with another ADMM based technique adopted from \cite{zheng2015fully} for TS-1 and TS-3. The convergence of the ADMM approach is demonstrated using the maximum residual at the boundary bus for TS-1 and TS-3 (see Fig. \ref{ADMM_Rs}). The two distributed methods are also compared for solution quality and compute time in Table \ref{ADMM_TS13}. Although the ADMM-based method converges to the same solution as the proposed ENApp, it takes around 272 macro-iterations for the mid-sized IEEE 123-bus system (TS-1) and 756 macro-iterations for the large IEEE 8500-node feeder (TS-3). Whereas, the proposed method requires 4 and 17 macro-iterations for IEEE 123-bus (TS-1) and IEEE 8500-node (TS-3), respectively to converge to a solution of the same quality. The required compute time to converge is also very high for the ADMM-based method.

\subsection{Extension to a Three-Phase Unbalanced Radial Distribution Feeder}
To showcase the efficacy of the proposed ENApp D-OPF algorithm for the unbalanced distribution network, we simulated a 3-phase unbalanced IEEE-13 bus system (TS-5) using our proposed method (See Fig. \ref{IEEE_13_unb} in Appendix A.). The local nonlinear unbalanced OPF model (P4) is described in Appendix A. The network is assumed to be composed of 2 areas with 3 DERs at each area. The result of the 3-phase unbalanced distribution system is shown in Table \ref{unbalance_3p}. From the table, we can see that all the 6 DERs have the same optimal Q injections for both proposed ENApp D-OPF and an equivalent C-OPF algorithm. Thus, the proposed approach is able to obtain an optimal solution of the same quality as the C-OPF model for an unbalanced distribution feeder. 
Also, both methods achieve the same minimum value for the active power losses of are 44.5 kW. The ENApp D-OPF converges within 2 iterations for this unbalanced system. Thus, the proposed ENApp D-OPF is also as efficient as for single-phase systems.

\begin{table}[t]
\renewcommand{\arraystretch}{1.4}
\centering
\caption{C-OPF vs. ENApp D-OPF for Unbalanced System (TS-5)}
{\begin{tabular}{ |c| c|c|c|}
\hline
{}&{Bus}&{C-OPF$^{\dag}$} &{D-OPF$^{\dag}$}\\
\hline
\parbox[t]{3mm}{\multirow{6}{*}{\rotatebox[origin=c]{90}{Q Injection}}}&{3 (A)}&{66.3 kVAr}&{66.3 kVAr}\\\cline{2-4}
{}&{3 (B)}&{66.3 kVAr}&{66.3 kVAr}\\\cline{2-4}
{}&{3 (C)}&{66.3 kVAr}&{66.3 kVAr}\\\cline{2-4}
{}&{7 (A)}&{132.7 kVAr}&{132.7 kVAr}\\\cline{2-4}
{}&{7 (B)}&{132.7 kVAr}&{132.7 kVAr}\\\cline{2-4}
{}&{7 (C)}&{132.7 kVAr}&{132.7 kVAr}\\\cline{1-4}
\multicolumn{2}{|c|}{Total Loss}&{44.5 kW}&{44.5 kW}\\\cline{1-4}
\multicolumn{3}{|c|}{Macro-Iteration for D-OPF$^{\dag}$}&{2}\\
\hline
\end{tabular}}
\label{unbalance_3p}
\vspace{-0.5cm}
\end{table}
}

  \vspace{-0.2cm}
\subsection{Discussions}
The advantages of the proposed ENApp D-OPF approach are briefly summarized here. First, the fewer rounds of communication required among neighbors in the proposed D-OPF model make it practically feasible and relatively less expensive to deploy than popular D-OPF methods (that take several hundred macro-iterations). Besides, the fewer rounds of communication among neighboring areas introduce lesser and manageable communication delays. The sub-linear increase in the iteration count with the increase in the number of the decomposed area also bodes well for scaling the approach for any size of the distribution system. Further, the proposed method does not require any parameter tuning to achieve convergence. If the original problem has the convergence guarantees, then the distributed version is also guaranteed to converge. This statement is also true for nonlinear optimization problems. Moreover, lower observability requirements improve overall system security. Finally, the proposed method is easily scalable to larger distribution systems without losing any of the advantages related to convergence. 

{\color{black} Next, we briefly detail the limitations of the proposed algorithm. As detailed before, the proposed approach actively leverages the radial topology of the distribution feeder. Thus, the developed algorithm, in its current form, is not applicable to general meshed feeders. A particular case of meshed topology, where each area is radially connected while sub-areas are meshed, can be solved using the proposed approach. The expansion of the proposed methods for weekly meshed systems is a part of future work.}

{\color{black}Finally, it is to be noted that advanced communication infrastructure is required to implement the proposed D-OPF method for the distribution systems. Although the current distribution feeders do not support such advanced communication capabilities, moving forward with the integration of ADMS and push from the utilities to better manage their distribution systems and distributed energy resources (DERs), it is only a matter of time when such information will be available and can be used in everyday optimization of the distribution systems \cite{Ngo2020Investing,Boardman2020Advanced}.}

 \vspace{-0.5cm}
\section{Conclusions}
This paper presents a novel distributed optimization algorithm to solve the optimal power flow (OPF) problem for a radial distribution system. The proposed approach is based on the fundamental properties of power flow in radial power distribution systems that allow us to represent the decomposed radial distribution systems as equivalent voltage sources (for the upstream node) and aggregated loads (for the downstream nodes). We term this as the equivalent network approximation based distributed OPF (ENApp D-OPF) algorithm. Upon leveraging this property, the number of macro-iterations/communication-rounds required for the convergence of the distributed algorithm is significantly reduced. The proposed approach is validated using simulations on the single-phase equivalent of IEEE 123-bus and IEEE 8500-node test systems and shown to converge to the same optimal solutions as obtained by the centralized OPF (C-OPF) method. Further, for the IEEE 8500-node test system, while the nonlinear model for the C-OPF cannot be solved using commercial solvers, the proposed D-OPF algorithm solves the nonlinear programming problem and converges to the optimum solution within a reasonable time. {\color{black} The proposed method's applicability for a radial unbalanced three-phase system is also demonstrated using a small unbalanced distribution feeder. The extension of the proposed method for weakly meshed topologies is currently under investigation and is a part of our future research.}

{\color{black}
\appendices
\section{Unbalanced 3-phase Distributed OPF Model}
This section details the D-OPF formulation for an unbalanced power distribution system. The power flow equations are based on our prior work on unbalanced D-OPF \cite{jha2019bi}. A part of radial power distribution network (Area $A_m$) of $n$ buses is considered where, $\N_m = \{1,2,…,n\}$ denotes the set of buses in that system. Assume $\E_m$ denotes set of edges identifying distribution lines that connect the ordered pair of buses $(i,j)$, $\forall$ $i,j \in \N_m$. The three-phases of the system are denoted by $\{a,b,c\}$ and $\phi_i$ denotes the set of phases in the bus $i$. Let $v_j^p =|V_j^p|^2$ be the squared magnitude of voltage at bus $j$ for phase $p\in \phi_j$. Define $\phi_{ij} = \{pq : p\in \phi_i$ and $q\in \phi_j$  $\forall (i,j)\in \E\} $. Let $l_{ij}^{pq} =(|I_{ij}^p||I_{ij}^q|)$ be the squared magnitude of the line current flowing in the phase ${pq}\in \phi_{ij}$ of line $(i,j)$. Also, $S^{pq}_{ij}=P^{pq}_{ij}+ Q^{pq}_{ij}$ and $z^{pq}_{ij}=r^{pq}_{ij}+j x^{pq}_{ij}$, where $pq \in \phi_{ij}$. $S_{Lj}^p = p_{Lj}^p+jq_{Lj}^p$  denote the load connected at node $j$ of phase $p\in \phi_i$. $\delta^{pq}_{ij}$ is the angle difference between the phase currents. $\mathbb{Re}$ denotes the real part of the complex number. For detail on power flow model please refer to \cite{jha2019bi}. The OPF problem is detailed next. The representative IEEE 13-bus feeder decomposed into two areas is given in Fig. \ref{IEEE_13_unb}.

\begin{equation}
\small
\nonumber \text{(P4)}  \hspace{0.6cm} \text{Minimize} \sum_{p \in \phi_j, j:i \rightarrow j}{l^{p}_{ij}r^{p}_{ij}}	\end{equation}
\begin{flalign}\label{unb_opf}
\nonumber \text{s.t.}  \hspace{0.2cm} &P_{ij}^{pp} - \sum_{q \in \phi_j}{l_{ij}^{pq} \left(r_{ij}^{pq} \cos(\delta_{ij}^{pq})- x_{ij}^{pq} \sin(\delta_{ij}^{pq})\right)} \\ 
\nonumber  &=\sum_{k:j \rightarrow k}P_{jk}^{pp} + p_{Lj}^p- p_{Dj}^p \hspace{3.2cm}\\
\nonumber  &Q_{ij}^{pp} - \sum_{q \in \phi_j}{l_{ij}^{pq}\left(x_{ij}^{pq} \cos(\delta_{ij}^{pq})+ r_{ij}^{pq} \sin(\delta_{ij}^{pq})\right)}\\
\nonumber &= \sum_{k:j \rightarrow k}Q_{jk}^{pp}+q_{Lj}^p  - q_{Dj}^p - q_{Cj}^p \hspace{3.2cm}\\
\nonumber &v_j^p = v_i^p - \sum_{q \in \phi_j}{2 \mathbb{Re}\left[S_{ij}^{pq} (z_{ij}^{pq})^*\right]} + \sum_{q \in \phi_j}{z_{ij}^{pq} l_{ij}^{qq}}\\
\nonumber &+\sum_{q1,q2 \in \phi_j, q1 \neq q2}{2\mathbb{Re}\left[ z_{ij}^{pq1} l_{ij}^{q1q2}\left(\angle(\delta_{ij}^{q1q2})\right)(z_{ij}^{pq2})^*\right]}\\
\nonumber &(P_{ij}^{pp})^2 + (Q_{ij}^{pp})^2 = v_i^p  l_{ij}^{pp}\\
\nonumber &(l_{ij}^{pq})^2 = l_{ij}^{pp}  l_{ij}^{qq}\\
\nonumber &q_{Dj}^p \leq \sqrt{(s_{DR}^{p})^2 - (p_{Dj}^p)^2} \\
\nonumber &q_{Dj}^p \geq -\sqrt{(s_{DR}^{p})^2 - (p_{Dj}^p)^2} \\
\nonumber &\underline{V}^2\leq v_j^p \leq \overline{V}^2
 \end{flalign}

\begin{figure}[h]
\centering
\includegraphics[width=0.25\textwidth]{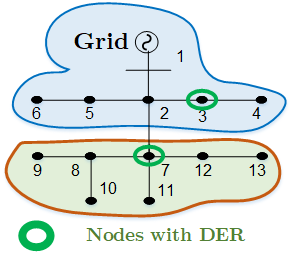}%
\vspace{-0.2cm}
\caption{\centering{TS-5: IEEE 13-bus test system with two areas.}}
\label{IEEE_13_unb}
\end{figure}

}


%





\ifCLASSOPTIONcaptionsoff
  \newpage
\fi

\bibliographystyle{ieeetr}
\bibliography{cite}
\end{document}